\documentclass[10pt]{article} 
\usepackage{spconf}
\usepackage{times}
\usepackage{amsmath,amssymb,float,arydshln,color}
\usepackage{psfrag,setspace,wrapfig,subfigure,float}
\usepackage[utf8]{inputenc}
\usepackage[font=small,labelfont=bf]{caption}
\usepackage{pifont}
\usepackage{dsfont}
\usepackage{epsfig}
\usepackage{epstopdf}
\usepackage{graphicx}
\usepackage{scrextend}
\usepackage{hyperref}
\usepackage[ruled]{algorithm2e}
\usepackage{amsfonts}
\usepackage{cite}
\usepackage{hyperref}
\usepackage{url}
\usepackage{booktabs}       
\usepackage{hhline}

\usepackage[table]{xcolor}
\usepackage{multirow}
\usepackage{tabu}
\allowdisplaybreaks

\newtheorem{lemma}{{Lemma}}
\newtheorem{assumption}{{ Assumption}}
\newtheorem{theorem}{{Theorem}}
\newtheorem{corollary}{{Corollary}}
\newcommand{\HRule}{\rule{\linewidth}{0.5mm}}

\def\tran{^{\mathsf{T}}}

\def\one{\mathds{1}}

\DeclareMathOperator*{\argmin}{arg\,min}
\newcommand{\bp}{ \begin{proof}}
	\newcommand{\ep}{\end{proof} }

\newcommand{\Ex}{\mathbb{E}\hspace{0.05cm}}

\newcommand{\be}{\begin{equation}}
\newcommand{\ee}{\end{equation}}
\newcommand{\bqq}{\begin{eqnarray}}
\newcommand{\eqq}{\end{eqnarray}}
\newcommand{\bal}{\begin{align}}
\newcommand{\eal}{\end{align}}
\newcommand{\bqn}{\begin{eqnarray*}}
	\newcommand{\eqn}{\end{eqnarray*}}
\newcommand{\nn}{\nonumber}
\newcommand{\ba}{\left[ \begin{array}}
	\newcommand{\ea}{\\ \end{array} \right]}
\newcommand{\qd}{\hfill{$\blacksquare$}}
\newcommand{\define}{\;\stackrel{\Delta}{=}\;}

\def\bphi  {{\boldsymbol \phi}}

\def\Pr {{\mathbb{P}}}

\def\A{{\boldsymbol{A}}}
\def\B{{\boldsymbol{B}}}

\def\Q{{\boldsymbol{Q}}}

\def\a{{\boldsymbol{a}}}
\def\b{{\boldsymbol{b}}}

\def\n{{\boldsymbol{n}}}

\def\p{{\boldsymbol{p}}}

\def\r{{\boldsymbol{r}}}

\def\v{{\boldsymbol{v}}}
\def\w{{\boldsymbol{w}}}
\def\x{{\boldsymbol{x}}}

\def\z{{\boldsymbol{z}}}

\newcommand{\cN}{{\mathcal{N}}}

\newcommand{\cU}{{\mathcal{U}}}

\newcommand{\scp}{\boldsymbol{\scriptstyle{\mathcal{P}}}}
\newcommand{\sr}{{\scriptstyle{\mathcal{R}}}}

\newcommand{\sv}{\boldsymbol{\scriptstyle{\mathcal{V}}}}
\newcommand{\sw}{\boldsymbol{\scriptstyle{\mathcal{W}}}}
\newcommand{\sx}{\boldsymbol{\scriptstyle{\mathcal{X}}}}

\newcommand{\grad}{{\nabla}}

\newcommand{\eq}[1]{\begin{align}#1\end{align}}
\newcommand{\beqn}{\begin{eqnarray}}
\newcommand{\eeqn}{\end{eqnarray}}

\DeclareFontFamily{U}{mathx}{\hyphenchar\font45}
\DeclareFontShape{U}{mathx}{m}{n}{
	<5> <6> <7> <8> <9> <10>
	<10.95> <12> <14.4> <17.28> <20.74> <24.88>
	mathx10
}{}
\DeclareSymbolFont{mathx}{U}{mathx}{m}{n}
\DeclareFontSubstitution{U}{mathx}{m}{n}
\DeclareMathAccent{\widebar}{0}{mathx}{"73}

\def\real{{\mathbb{R}}}

\def\Zint{{\mathchoice{\setbox1=\hbox{\sf Z}\copy1\kern-.75\wd1\box1}
		{\setbox1=\hbox{\sf Z}\copy1\kern-.75\wd1\box1}
		{\setbox1=\hbox{\scriptsize\sf Z}\copy1\kern-.75\wd1\box1}
		{\setbox1=\hbox{\scriptsize\sf Z}\copy1\kern-.75\wd1\box1}}}
	
\title{Dynamic Average Diffusion with randomized Coordinate Updates}
\name{Bicheng Ying${}^{*\dagger}$, Kun Yuan${}^{*\dagger}$, and Ali H. Sayed${}^\dagger$\vspace{-0mm}\thanks{This work was supported in part by NSF grant CCF-1524250. Authors B. Ying and K. Yuan are with the Electrical Engineering Department, UCLA. A. H. Sayed is with the School of Engineering, EPFL, Switzerland. Emails: \{ybc, kunyuan\}@ucla.edu and ali.sayed@epfl.ch}}

\address{\normalsize{${}^*$Department of Electrical Engineering,
	University of California, Los Angeles}\\
\normalsize${}^\dagger$School of Engineering, 
\'Ecole Polytechnique F\'ed\'erale de Lausanne, Switzerland \vspace{0mm}}

\begin{document}
\def\helvetica{phvr7t.tfm}
\def\helveticaoblique{phvro7t.tfm}
\def\helveticabold{phvb7t.tfm}
\def\helveticaboldoblique{phvbo7t.tfm}

\font\sfb=\helveticabold
=\helveticaboldoblique
\maketitle
\small
\begin{abstract}\vspace{-0mm}
This work derives and analyzes an online learning strategy for tracking the average of time-varying distributed signals by relying on randomized coordinate-descent updates. During each iteration, each agent selects or observes a random entry of the observation vector, and different agents may select different entries of their observations before engaging in a consultation step. Careful coordination of the interactions among agents is necessary to avoid bias and ensure convergence. We provide a convergence analysis for the proposed methods, and illustrate the results by means of simulations. 
\end{abstract}\vspace{0mm}
\begin{keywords}
	dynamic average diffusion, consensus, push-sum algorithm, coordinate descent, exact diffusion.\vspace{-0mm}
\end{keywords}
\setlength{\abovedisplayskip}{1.2mm}
\setlength{\belowdisplayskip}{1.2mm}
\def\arraystretch{0.9}\vspace{-1mm}
\section{INTRODUCTION AND MOTIVATION}\vspace{0mm}
We consider the problem in which a collection of $K$ networked agents, indexed $k=1,2,\ldots,K$,  is interested in tracking the {\em average}  of time-varying signals $\{r_{k,i}\}$ arriving at the agents, where $k$ is the agent index and $i$ is the time index. The objective is for the agents to attain tracking in a decentralized manner through  local interactions with their neighbors. This type of problem is common in many applications.
 For example, consider the following distributed empirical risk minimization problem\cite{sayed2014adaptation, nedic2017achieving,chen2013distributed,kar2009distributed,shi2015proximal,yuan2017exact1,nedic2009distributed}, which arises in many traditional machine learning formulations:
\eq{
	w^\star = \min_{w\in\real^{M}} J(w) \define \frac{1}{K}\sum_{k=1}^K  Q(w;X_{k}) \label{prob.1}
}
where $Q(w;X_{k})$ is some loss function that depends on the data $X_k$ at location or agent $k$. 
If we let $w_{k,i}$ denote an estimate for the minimizer $w^{\star}$  at agent $k$ at time $i$, and let $X_{k,i}$ denote the data received at that agent at the same time instant, then some solution methods to (\ref{prob.1}) involve tracking the average gradient defined by\cite{nedic2017achieving,yuan2017exact1,xin2018linear}:
\eq{
	\bar{r}_{i} =	\frac{1}{K}\sum_{k=1}^K  \nabla_w Q(w_{k,i};X_{k,i})
}
where each term inside the summation represents the signal $r_{k,i}$. Likewise, in learning problem formulations involving feature vectors and parameter models that are distributed over space, or loss functions that are expressed in the form of sums \cite{chen2015dictionary, sundhar2012new, mota2012distributed,ying2017diffusion,ying2018exponentially, ying2018learning}, we encounter optimization problems of the form
\eq{
		w^\star = \min_{w\in\real^{M}} J(w) =  \frac{1}{N}\sum_{n=1}^NQ\Big(\frac{1}{K}\sum_{k=1}^K f(w_k;X_{k,n})\Big) \label{prob.2}
}
where $f(w_k;X_{k,n})$ is some linear or nonlinear function that depends on the $n-$th feature set, $X_{k,n}$, available at agent $k$. Some solution methods to (\ref{prob.2}) involve tracking the average quantity:
\eq{
\bar{r}_{i} =	\frac{1}{K}\sum_{k=1}^K  \ba{c}   f(w_{k,i};X_{k,1})\\[-1mm]\vdots\\ f(w_{k,i};X_{k,N})\ea
}
where again each term inside the summation represents an $r_{k,i}$ signal. 

There are several useful distributed algorithms in the literature for computing the average of {\em static} signals $\{r_k\}$ (i.e., signals that do not vary with the time index $i$), and which  are distributed across a network \cite{kar2011convergence, chen2013distributed,sayed2014adaptation,nedic2009distributed,shi2015extra, boyd2006randomized}. One famous algorithm is the consensus strategy which takes the form
\eq{
	w_{k, i} = \sum_{\ell \in \cN_k} a_{\ell k} w_{\ell,i-1},\;\;\;{\rm where\ } w_{k, 0} = r_k \label{static.consensus}
}
where $a_{\ell k}$ is a nonnegative factor scaling the information from agent $\ell$ to agent $k$ and $A=[a_{\ell k}]$  is some doubly-stochastic matrix. Moreover, the notation ${\cal N}_k$ denotes the set of neighbors of agent $k$. In this implementation, each agent starts from its observation vector $r_k$ and continually averages the state values of its neighbors. After sufficient  iterations, it is well-known that 
\eq{
	w_{k, i}\to \frac{1}{K}\sum_{k=1}^K r_{k} \label{consensus}
} under some mild conditions on $A$ \cite{sayed2014adaptive,boyd2006randomized,horn1990matrix,pillai2005perron, nedic2009distributed}. When the static signals $\{r_k\}$ become dynamic and are replaced by $\{r_{k,i}\}$, a useful variation is the dynamic average consensus algorithm from \cite{freeman2006stability, zhu2010discrete, cao2013overview}. It replaces (\ref{static.consensus}) by the recursion:
\eq{
	w_{k, i} = \!\sum_{\ell \in \cN_k} \!\!a_{\ell k} w_{\ell, i-1}\! + \!r_{k, i}\! -\! r_{k, i-1},\;\;{\rm where\ } w_{k, 0} = r_{k,0} \label{fwefkop}
}
where the difference $r_{k,i}-r_{k,i-1}$ is added as a driving term. In this case, it can be shown that if the signals $\{r_{k,i}\}$ converge to static values, i.e., if $r_{k,i}\rightarrow r_k$, then result \eqref{consensus} continues to hold \cite{freeman2006stability, zhu2010discrete}. Recursion (\ref{fwefkop}) is motivated in \cite{freeman2006stability, zhu2010discrete} using useful but heuristic arguments.  

Motivated by these considerations, in this work, we develop a dynamic average diffusion strategy for tracking the average of time-varying signals $\{r_{k,i}\}$ by formulating an optimization problem and showing how to solve it by applying the exact diffusion strategy from \cite{yuan2017exact1,yuan2017exact2}. {\color{black}  One of the main contributions relative to earlier approaches is that we are specifically interested in the case in which the dimension of the observation vectors $\{r_{k,i}\}$ may be too large, which means that a solution like (\ref{fwefkop}) will necessitate the sharing of long vectors $w_{k,i}$ among the agents resulting in an inefficient communication scheme, especially in real-time processing scenarios. We are also interested in the case in which each agent $k$ can only observe one random entry of $r_{k,i}$ at each iteration (either by design or by choice). 
In this case, it will be wasteful to share the full vector $w_{k,i}$ since only one entry of $w_{k,i}$ will be affected by the new information. To handle these situations, 
we will need to incorporate elements of randomized coordinate-descent\cite{tseng2001convergence, luo1992convergence, nesterov2012efficiency, richtarik2014iteration} into the operation of the algorithm in line with approaches from \cite{arablouei2014distributed, defazio2014saga, wang2016coordinate}. Therefore, the motivation for choosing coordinate-wise updates is mainly due the communication and real-time processing constraints. Doing so, however, introduces one nontrivial complication: different agents may be selecting or observing different entries of their vectors $r_{k,i}$, which raises a question about how to coordinate or synchronize their interactions. \vspace{-0.5mm}

In order to facilitate the presentation, we shall assume initially that all agents select the {\em same}  entry of their observation vectors at each iteration. Subsequently, we will show how to employ push-sum ideas \cite{nedic2015distributed, nedic2016stochastic, benezit2010weighted, assran2018stochastic} to allow each agent to select its own local entry {\em independently} of the other agents. 
\vspace{-0.5mm}

{\bf Related works}: This paper combines three techniques (dynamic averaging, coordinate-wise updates, and the push-sum method) into a consolidated method for online tracking. Although there have been works in the literature on these techniques separately, they have been rarely combined into a unifying tool within the same framework. Here are some overviews of prior works:\\
 1) {\it Dynamic average algorithm}: The algorithm has been used before, e.g., in \cite{freeman2006stability,zhu2010discrete,hu2012robust}. A variety of exact first-order distributed algorithms\cite{shi2015extra, xin2018linear, nedic2017achieving, di2016next,yuan2017exact1,yuan2017exact2} have also been researched in recent years. However, to the best of our knowledge, this paper appears to be the first to combine these two types of methodologies to solve the online tracking problem.  \\
 2) {\it Coordinate-wise updates}: 
 There is extensive literature on using coordinate-descent updates for static signals\cite{tseng2001convergence, nesterov2012efficiency,richtarik2014iteration,zouzias2015randomized,peng2016arock}. We, however, focus on dynamic signals. One main challenge in the dynamic scenario is that one may not be able to retrieve the coordinate gradient or other coordinate-wise information from the previous iteration due to evolution over time. To address this challenge, we exploit the SAGA technique\cite{defazio2014saga,defazio2014finito} to introduce an auxiliary dynamic memory, which helps balance the requirement of dynamic signal tracking and reduced computations.  As a result, the paper also needs to account for the presence of delays, which sets the analysis apart from other works \cite{wang2016coordinate,hendrikx2018accelerated}.\\
	{\color{black} 3) {\it Push-sum method}: 
		This method corrects the bias in the algorithm due to asymmetry in the network topology \cite{kempe2003gossip}.  
	The push-sum method has been introduced in \cite{kempe2003gossip} to correct the bias when perform the consensus task in the asymmetry structure of network.
	 The main idea is to keep updating another set of weights along with the signal.
	 The construction has been extended from the pure consensus formulation to distributed optimization problems \cite{tsianos2012push, nedic2015distributed, benezit2010weighted, dominguez2011distributed,assran2018stochastic,notarnicola2018distributed}. Reference \cite{notarnicola2018distributed} has relations to the current manuscript in that they studied a block/coordinate algorithm with an embedded push-sum strategy. However, this reference focused on {\em static} distributed optimization problems. Moreover, the use of the push-sum strategy in our current manuscript is used to address the induced flow imbalance of information that is generated by the coordinate updates. Also, unlike earlier works on dynamic push-sum algorithms \cite{nedic2015distributed, notarnicola2018distributed, nedic2016stochastic}, which assume $B-$strongly connected networks, the analysis in the current work is based on a more relaxed unbounded assumption. 
	 
}

{\color{black}
	The references mentioned so far focus mainly on deriving solutions in the primal domain and use first-order distributed algorithms for tracking. There are of course other families of distributed methods for tracking dynamic signals, such as those based on the ADMM procedure \cite{simonetto2017decentralized, ling2013decentralized}, and the distributed Kalman filter \cite{olfati2007distributed, olfati2005distributed, cattivelli2010diffusion}.  These are powerful methods with good convergence rates. Nevertheless, these methods have high communication requirements among agents and the design of coordinate-wise techniques for them is more challenging. The main focus of our manuscript is on the development of a tracking method for scenarios with limited communication bandwidth and where the processing latency is critical. 
}

	{\bf Notation}: We use plain letters for
	deterministic variables, and boldface letters for random variables. We also use $\Ex_x$ to denote the expectation with respect to $x$, ${\rm col}\{x_1,\cdots,x_n\}$ to denote a column vector formed by stacking $x_1,\cdots,x_n$, $(\cdot)\tran$ to denote transposition, $\odot$ to Hadamard  production, and $\|\cdot\|$ for
	the 2-norm of a matrix or the Euclidean norm of a vector.
	Throughout the paper, we use the subscripts $i, j$ to index iterations and time, and $k, \ell$ as the index of agent, $w(n)$ to refer to the $n$-th entry of vector $w$. We also sometimes use $i, j$ as superscripts to refer to iterations and time. The notation $\one_N= {\rm col}\{1,\ldots,1\}\in\real^N$ and $\cN_k$ represents the neighborhood set of agent $k$. 
}

%

{\color{black}
\section{MOTIVATION OF BASE ALGORITHM}
In this section, we derive a basic distributed tracking strategy referred to as  dynamic average diffusion, which will serve as the cornerstone for our later discussion and the more general algorithm derivation shown further ahead. Dynamic average diffusion is similar to dynamic average consensus \eqref{fwefkop} except that the combination matrix $A$ is also applied to the dynamic signal $r_k$ -- see \eqref{dynamic.diffusion} below. We shall derive the dynamic average diffusion strategy by adapting the derivation of the exact diffusion method \cite{yuan2017exact1,yuan2017exact2}  to the dynamic signal tracking scenario. To do so, we will need to introduce an appropriate cost function. This is approach is in contrast to the derivation of the dynamic average consensus method, which has a close relationship to the bias-corrected method known as EXTRA \cite{shi2015extra} but does not actually follow from it directly.
In this section, we also derive the dynamic tracking methods based on other popular distributed optimization strategies such as gradient-tracking\cite{xu2015augmented, xin2018linear, nedic2017achieving,notarnicola2018distributed,di2016next, xi2017add, qu2017harnessing} and compare them with the proposed dynamic average diffusion approach.

}

\subsection{Review of Exact Diffusion Strategy}\vspace{-2mm}
One effective decentralized method to solve problems of the form:
\eq{
	w^\star \define \argmin_{w\in\real^{M}} J(w) \define \frac{1}{K}\sum_{k=1}^{K} J_k(w) \label{prob-emp-dist}
} is the Exact diffusion strategy \cite{yuan2017exact1,yuan2017exact2}. In (\ref{prob-emp-dist}), each $J_k(w)$ refers to the risk function at agent $k$ and is generally convex or strongly-convex. For simplicity, we shall assume in this work that each $J_k(w)$ is differentiable although the analysis can be extended to non-smooth risk functions by employing subgradient constructions, along the lines of \cite{nedic2009distributed, ying2018performance}, or proximal constructions similar to \cite{vlaski2016diffusion,shi2015proximal}. To implement exact diffusion, we need to associate a combination matrix $A=[a_{\ell k}]_{\ell,k=1}^K$ with the network graph, where a positive weight $a_{\ell k}$ is used to scale data that flows from node $\ell$ to $k$ if both nodes happen to be neighbors.  In this paper we assume that:\vspace{-1mm}
\begin{assumption}[\sc Topology]\label{assumption-topology} The underlying topology is strongly connected, and the combination matrix 
	$A$ is symmetric and doubly stochastic, i.e.,\vspace{-0.5mm}
	\eq{
		A = A\tran\;\mbox{and}\;\; A \mathds{1}_K = \mathds{1}_K
	}
	where $\mathds{1}$ is a vector with all unit entries. {\color{black} We further assume that $a_{kk}>0$ for at least one agent $k$.} 
	$\hfill\Box$
\end{assumption}
We further introduce $\mu$ as the positive  step-size parameter for all nodes. The exact diffusion algorithm is listed in (\ref{ed-adapt})--(\ref{ed-comb}). It is shown in \cite{yuan2017exact2} that the local variables $w_{k,i}$ converge to the exact minimizer of problem \eqref{prob-emp-dist}, $w^\star$,  at a linear convergence rate under relatively mild conditions.

\begin{table}[!h]
	\vspace{1mm}
	\noindent \HRule\\
	\noindent \textbf{\small Algorithm 1 [Exact diffusion strategy for each node $k$]}\cite{yuan2017exact1,yuan2017exact2} \vspace{-2mm}\\
	\HRule\\
	{\bf Initialize} $w_{k,0}$ arbitrarily, and let $ \psi_{k,0} = w_{k,0}$.\\
	\textbf{{ Repeat}} iteration $i=1,2,3\cdots$\vspace{-2mm} \textbf{until convergence}
	{\begin{subequations}
		\eq{
			\hspace{-10mm}\psi_{k,i} &= w_{k,i-1} - \mu\, \grad_w J_k(w_{k,i-1}) \label{ed-adapt}\\
			\hspace{-10mm}\phi_{k,i} &= \psi_{k,i} + w_{k,i-1}  - \psi_{k,i-1} \label{ed-correct} \\
			\hspace{-10mm}w_{k,i} &= \sum_{\ell\in \cN_k} {a}_{\ell k} \phi_{\ell,i} \label{ed-comb}\\[-7.5mm]\nn
		}\vspace{1mm}
	\end{subequations}
	}\\
	\HRule\vspace{-4mm}
\end{table}

\subsection{Dynamic Average Diffusion}\label{sec-dad}
Now, we consider a time-varying quadratic risk  function of the form
\be
J_{k,i}(w)=\frac{1}{2}\|w-r_{k,i}\|^2
\ee
and introduce the average cost
\be
J_i(w)\define \frac{1}{K}\sum_{k=1}^K J_{k,i}(w)       \label{xdsfew.d}
\ee
At every time instant $i$, if we optimize $J_i(w)$ over $w$ then it is clear that the minimizer, denoted by $w^o_i,$ will coincide with the average of the observed signals:
\be
w^o_i=\bar{r}_i\define \frac{1}{K}\sum_{k=1}^K r_{k,i}
\ee
Therefore, one way to track the average of the signals $\{r_{k,i}\}$ is to track the minimizer of the aggregate cost $J_i(w)$ defined by (\ref{xdsfew.d}). Apart from the time index, this cost has a form similar to (\ref{prob-emp-dist}) especially when the observations signals $\{r_{k,i}\}$ approach steady-state values where they become static. This motivates us to apply the exact diffusion construction (\ref{ed-adapt})--(\ref{ed-comb}) to the risks defined by \eqref{xdsfew.d}. Doing so leads to the recursions:
\begin{subequations}
	\eq{
		\psi_{k,i} 	=&\;  (1-\mu)w_{k,i-1} +\mu r_{k,i}\label{exact.diffusion1}\\
		\phi_{k,i} =&\; \psi_{k,i} + w_{k,i-1} -  \psi_{k,i-1}\label{exact.diffusion2}\\
		w_{k,i} =&\; \sum_{\ell \in \cN_k}a_{\ell k} \phi_{\ell,i}\label{exact.diffusion3}
	}
\end{subequations}
Combining (\ref{exact.diffusion1})--\eqref{exact.diffusion3} into a single recursion, we obtain:
\eq{
	w_{k,i} =&\; \sum_{\ell \in \cN_k}a_{\ell k}\Big((1-\mu)w_{\ell,i-1} +\mu  r_{\ell,i} + w_{\ell,i-1}\nn\\[-2mm]
	&\hspace{17mm} - (1-\mu)w_{\ell, i-2} -\mu  r_{\ell,i-1}  \Big)\label{23g2s}
}
so that by selecting $\mu=1$, the algorithm reduces to what we shall refer to as the dynamic average diffusion algorithm:
\begin{table}[h]
	\vspace{1mm}
	\noindent \HRule\\
	\noindent \textbf{\small Algorithm 2 [Dynamic average diffusion]} \vspace{-2mm}\\
	\HRule\\
	{\bf Initialize}: $w_0=r_{k,0}$.\\
	\textbf{{ Repeat}} iteration $i=1,2,3,\ldots$\textbf{until converge}\vspace{-2mm}
	{
		\eq{
			w_{k,i} = \sum_{\ell \in \cN_k}a_{\ell k}(w_{\ell,i-1}+r_{\ell,i} - r_{\ell,i-1})\label{dynamic.diffusion}\\[-10mm]\nn
		}
	}
	\HRule\vspace{-2mm}
\end{table}

 
 Other values for $\mu$ are of course possible by using instead (\ref{23g2s}). Comparing (\ref{dynamic.diffusion}) with  the consensus version (\ref{static.consensus}), we see that the scaling weights $\{a_{\ell k}\}$  in \eqref{dynamic.diffusion} are multiplying  the combined sum of the weight iterate $w_{\ell,i-1}$ and the difference of the current and past observation vectors, $r_{\ell,i}-r_{\ell,i-1}$. Moreover, and importantly, while in the consensus construction  (\ref{static.consensus}) each agent $k$ employs only its own observation vector, we see in (\ref{dynamic.diffusion}) that all observations vectors from the neighborhood ${\cal N}_k$ of agent $k$ contribute to the update of $w_{k,i}$. In this way, agents need to share their weight iterates along with the difference of their observation vectors. In a future section, we shall show how agents can only share single entries of their observations vectors chosen at random.  

There are several interesting properties associated with the dynamic diffusion strategy \eqref{dynamic.diffusion}.
First, at any time $i$, the average of the $\{w_{k,i}\}$ coincides with the average of the $\{r_{k,i}\}$, i.e.,
\eq{
	\frac{1}{K}\sum_{k=1}^K w_{k,i} = \frac{1}{K}\sum_{k=1}^Kr_{k,i} ,\;\;\;\;\forall i
	\label{eq.same.mean}
}
This property can be easily shown using mathematical induction. Second, when the signal is static, i.e., $r_{k,i} \equiv r_{k}$, {\color{black} Algorithm 2} reduces to the classical consensus construction (\ref{static.consensus}). Third, when the signal $r_{k,i}$ converges to some steady-state value $r_k$, or their time variations become uniform after some time $i_o$, i.e., 
\eq{
	r_{k,i} - r_{k,i-1} = r_{k',i} - r_{k',i-1},\;\;\; \forall k,k', i>i_o
}
then it can be shown that  
\eq{
	\lim_{i\to\infty} \|w_{k,i} - \bar{r}_{i} \| = 0   
}
This conclusion is a special case of later results in this paper and therefore its proof will follow by  specializing the arguments used later in {\color{black}Theorem \ref{theorem.1}}. 

{\color{black}\noindent{\bf Remark:} The setting  we consider in this paper  is the dynamic and continuous tracking scenario, which typically runs the algorithm continually or stops after some specified number of iterations. However, in some cases, it can be useful to employ a stopping criterion, for example, when the dynamic signal becomes static after some stages and there is no need to continue tracking it.   One feasible policy is for each agent to employ the comparison:
	\eq{
		\|w_{k,i} - w_{k,i-1}\| \leq \epsilon	 \label{jiog.weg}		
	}
	It is crucial to note that different agents might satisfy \eqref{jiog.weg} at different iterations, i.e., some agents might satisfy the stopping criterion earlier than other agents.  Therefore, when agents meet their convergence criteria, they can stop updating their iterates $w_{k,i}$ but continue to respond to communication requests from their neighbors. 
}

{\color{black}
\subsection{Derivation based on other methods}\label{sec.other.algorithm}
We could also attempt to apply similar arguments to other distributed algorithms in an effort to obtain other variations for dynamic averaging. However, as the analysis will show, these other methods will lead to more complex solutions than what is proposed in (\ref{dynamic.diffusion}). For example, if we apply the EXTRA algorithm\cite{shi2015extra} to \eqref{xdsfew.d}, we have
\eq{
	w_{k,i} =& \sum_{\ell \in \cN_k}\widetilde{a}_{\ell k} w_{\ell,i-1} - \sum_{\ell \in \cN_k}\frac{\widetilde{a}_{\ell k}}{2} w_{i-2,\ell}\nn\\
	& \;- \mu \Big( \nabla_w J_{k,i}(w_{k,i-1}) - \nabla_w J_{k,i-1}(w_{i-2,k})\Big)\nn\\
	=& \sum_{\ell \in \cN_k}\widetilde{a}_{\ell k} w_{\ell,i-1} - \sum_{\ell \in \cN_k}\frac{\widetilde{a}_{\ell k}}{2} w_{i-2,\ell}\nn\\
	& \;- \mu\Big( w_{k,i-1}-r_{k,i} -  w_{i-2,k} + r_{k,i-1} \Big)
}
where 
\eq{
	\widetilde{a}_{\ell k} = \begin{cases}
		a_{\ell k}&\ell\neq k\\
		1+a_{\ell k} & \ell= k
	\end{cases}
}
Similarly, after setting $\mu=1$, we have
\eq{
	w_{k,i} =& \sum_{\ell \in \cN_k}a_{\ell k} w_{\ell,i-1}+  r_{k,i} - r_{k,i-1}  - \sum_{\ell \in \cN_k}\frac{\widetilde{a}_{\ell k}}{2}w_{i-2,\ell} + w_{i-2,k} \label{extra-based.alg}
}
Compared with dynamic average consensus:
\eq{
		w_{k,i} =& \sum_{\ell \in \cN_k}a_{\ell k} w_{\ell,i-1}+  r_{k,i} - r_{k,i-1} 
		\label{consensus-based.alg}
}
we see that we have an extra term:
\eq{
	-\sum_{\ell \in \cN_k}\frac{\widetilde{a}_{\ell k}}{2}w_{i-2,\ell} + w_{i-2,k}
}
Although it will be zero as $w_{k,i}$ reaches consensus, it increases the algorithm complexity unnecessarily for this problem \eqref{xdsfew.d} due to the asymmetric structure.

Another common approach is gradient-tracking based algorithms\cite{xu2015augmented, xin2018linear, nedic2017achieving,notarnicola2018distributed,di2016next, xi2017add, qu2017harnessing}. Taking DIGing\cite{nedic2017achieving} algorithm as example:
\eq{
	\begin{cases}
		w_{k,i} =& \displaystyle\sum_{\ell\in\cN_k}a_{\ell k} \left(w_{\ell,i-1} - \mu y_{\ell,i-1}\right)\\
		y_{k,i} =& \displaystyle\sum_{\ell\in\cN_k}a_{\ell k} \left(y_{\ell,i-1} + \nabla_w J_{\ell,i}(w_{\ell,i}) - \nabla_w J_{\ell,i-1}(w_{\ell, i-1})\right)
	\end{cases}
}
Using a similar argument and setting $\mu=1$, we have
\eq{\label{2378sdbsd9}
	\begin{cases}
		w_{k,i} =& \displaystyle\sum_{\ell\in\cN_k}a_{\ell k}\left( w_{\ell,i-1} -  y_{\ell,i-1}\right)\\
		y_{k,i} =& \displaystyle\sum_{\ell\in\cN_k}a_{\ell k} \Big[y_{\ell,i-1} + \Big( w_{\ell,i}-r_{\ell,i} -  w_{\ell, i-1} + r_{\ell,i-1} \Big)\Big]
	\end{cases}
}
Note that recursion \eqref{2378sdbsd9} requires two rounds of communication, which is a natural issue stemming from gradient-tracking based algorithms --- one communication is to combine the primary variable $w$ and the other is for the gradient information. 

{\color{black} The performance comparison between consensus, diffusion, EXTRA, exact diffusion and gradient-tracking on the {\em  static} cost fucntion are well studied in \cite{tu2012diffusion,yuan2017exact2,nedic2017achieving,li2017decentralized}. We will compare all above derived methods in the dynamic scenario later in Sec. \ref{sec.simulation} --- see Fig. \ref{fig.diff_algs}. 
In following sections, we will extend dynamic average diffusion to the random coordinate update case in order to exchange only one coordinate or one block of coordinates per iteration. It is worth pointing out that the technique we derive in the following sections is {\em not} tied to \eqref{dynamic.diffusion} although the  derivation will be based on \eqref{dynamic.diffusion}.
	
%
%
}
}

\section{Synchronized Random Updates} 
%
Let us consider next the case in which each agent $k$ can only access (either by design or by choice) {\em one}  random entry within the vector $r_{k,i}$. We denote the index of that entry by $\n_i$ at iteration $i$; we use the boldface notation because $\n_i$ will be selected at random and boldface symbols denote random quantities in our notation.  We shall first assume that all agents select the {\em same} $\n_i$; later we consider the case in which $\n_i$ varies among agents and replace the notation by $\n_i^k$ instead, with the superscript $k$ referring to the agent. This situation will then enable a fully distributed solution. 

When all agents select the same random index $\n_i$, one naive solution to update their weight iterates is to resort to coordinate-descent type constructions\cite{luo1992convergence,tseng2001convergence}. Namely, at iteration $i$, the index $\n_i$ is selected uniformly and then only the $\n_i-$th entry of $\w_{k,i}$ is updated, say, as:
\eq{\everymath={\displaystyle}
\begin{cases}
	\w_{k,i}(\n_i)&\hspace{-3mm}
	= \sum_{\ell \in \cN_k}a_{\ell k} \big(\w_{\ell,i-1}(\n_i)+r_{\ell,i}(\n_i) - \underbrace{r_{\ell,i-1}(\n_i)}_{\rm \color{black}unavailable}\big)\label{238ds}\\
	\w_{k,i}(n)
	&\hspace{-3mm}= \w_{k,i-1}(n),\;\;\;n \neq \n_i
\end{cases}\raisetag{4mm}
}
where the notation $w(n)$, for a vector $w$, refers to the $n-$th entry of that vector. This iteration applies (19) to the $\n_i-$th entry of $\w_{k,i}$ and keeps  all other entries of this vector unchanged relative to $\w_{k,i-1}$. Although simple, this algorithm is {\em not} implementable for one subtle reason.
This is because at time ${i-1}$, agent $\ell$ can only observe $r_{\ell,i-1}(\n_{i-1})$ and not $r_{\ell,i-1}(\n_i)$. In other words, the variable  $r_{\ell,i-1}(\n_i)$ is not available; this variable would be available if we allow agent $\ell$ to save the entire vector $r_{\ell,i-1}$ from the previous iteration and then select its $\n_i-$th entry at time $i$. However, doing so, defeats the purpose of a coordinate-descent solution where the purpose is to avoid  working with long observation vectors and to work instead with scalar entries. We can circumvent this difficulty as follows. We let $j$ refer to the most recent iteration from the past where the {\em same} index $\n_i$ was chosen  the last time; the value of $j$ clearly depends on $\n_i$. Then, we can replace (\ref{238ds}) by:
\eq{\everymath={\displaystyle}
	\begin{cases}
\w_{k,i}(\n_i)&\hspace{-3mm}
= \sum_{\ell \in \cN_k}a_{\ell k} \big(\w_{\ell,j}(\n_i)+r_{\ell,i}(\n_i) - r_{\ell, j}(\n_i)\big)\label{g32g32d}\\
\w_{k,i}(n)&\hspace{-3mm}
= \w_{k,i-1}(n),\;\;\;n \neq \n_i
\end{cases}
} 
where the index $j$ appears in two locations on the right-hand side: within $\w_{\ell,j}$ and $r_{\ell,j}(\cdot)$. Note first that 
this implementation is now feasible because the scalar value $r_{\ell,j}(\n_i)$ from the past can be saved into a memory variable. Specifically, for every agent $k$ we introduce a vector $\v_{k,i}$, which is updated with time. At every iteration $i$, an index $\n_i$ is selected and the value of the observation entry $\r_{k,i}(\n_i)$ is saved into the $\n_i-$th location of $\v_{k,i}$ for later access the next time the index $\n_i$ is selected.  
It is also important to use $\w_{\ell,j}(\n_i)$, with the same subscript $j$, along with $r_{\ell,i}(\n_i)$ in \eqref{g32g32d} in order to maintain the mean property \eqref{eq.same.mean}.
However, due the definition of $j$ and the second line in \eqref{g32g32d} , we know that $\w_{\ell,j}(\n_i) = \w_{\ell,i-1}(\n_i)$.
Hence, the resulting algorithm is:
\eq{\everymath={\displaystyle}
	\begin{cases}
		\w_{k,i}(\n_i)
		&\hspace{-3mm}= \sum_{\ell \in \cN_k}a_{\ell k} (\w_{\ell,i-1}(\n_i)+r_{\ell,i}(\n_i) - \v_{\ell,i-1}(\n_i))\label{algorithm1.1}\\
		\w_{k,i}(n)
		&\hspace{-3mm}= \w_{k,i-1}(n),\;\;\;n \neq \n_i\\
		\v_{k,i}(n) &\hspace{-3mm}=\begin{cases}
			\r_{k,i}(n),&\mbox{if } n=\n_i\\
			\v_{k,i-1}(n),&\mbox{if } n\neq\n_i
		\end{cases}
	\end{cases}\raisetag{8mm}
}
To simplify the notation, we introduce the indicator function:
\eq{
	\mathbb{I}[{\rm expression}] \define \begin{cases}
		1,&\mbox{if expression is true} \\
		0, &\mbox{if  expression is false} 
	\end{cases} \label{indicator}
}
and  the selection matrix:
\eq{
	\mathbb{S}_{\n_i} \define \ba{cccc}
	\mathbb{I}[\n_i=1]&&&\\
	&\mathbb{I}[\n_i=2]&&\\
	&&\ddots&\\
	&&&\mathbb{I}[\n_i=N]
	\ea\label{fvowain}
}
This matrix is diagonal with a single unit entry on the diagonal at the location of the active index $\n_i$. All other entries are zero. We also introduce the complement matrix: 
\eq{
	{\mathbb{S}}_{\n_i}^c\define I_{N} - \mathbb{S}_{\n_i}
}
Using these matrices, the resulting algorithm is listed in Algorithm 3. The proof of the convergence is provided later in Sec. \ref{sec.conv.alg2}.
\begin{table}[t]
	\noindent\HRule\\
	\textbf{Algorithm 3 [Dynamic average diffusion with synchronous random updates]}\\[-2mm]
	\noindent\HRule\\
	{\bf Initialization:} \mbox{set $w_{k,0}=r_{k, 0}$; $v_{k,0}=r_{k,0}.$} \\
	{\bf Repeat for $i=1,2,\ldots$ until convergence:}\vspace{-2mm}
	\begin{subequations}
		\begin{align}
		\n_i \sim&\, \cU[1,N] \hspace{10mm}\mbox{(uniform sampling)}\label{dd.1}\\
		\w_{k,i} =&\, {\mathbb{S}}_{\n_i}^c \w_{k,i-1} +  \sum_{\ell \in \cN_\ell}a_{ \ell k}\,\mathbb{S}_{\n_i}(\w_{\ell,i-1}+r_{\ell,i} - \v_{\ell,i-1})\label{dd.2}\raisetag{2mm}\\
		\v_{k,i} =&\, {\mathbb{S}}_{\n_i}^c\v_{k,i-1}+  \mathbb{S}_{\n_i}r_{k,i}\label{dd.3}\\[-9mm]\nn
		\end{align}
	\end{subequations}
	\noindent\HRule\vspace{-4mm}
\end{table}
%

\section{Independent Random Updates}
\subsection{A first attempt at random indices}
{\color{black} Algorithm 3 }requires  all agents to observe the same ``random" index $\n_i$ at iteration $i$. In this section, we will allow $\n_{i}$ to be locally selected by the agents. To refer to this generality, we replace the notation $\n_i$ by $\n_i^k$, where   $\n_{i}^k$ is selected uniformly from 
$\{1,2,\ldots,N\}$ by agent $k$.

\begin{figure}[h]
	\centering
	\includegraphics[scale=0.38]{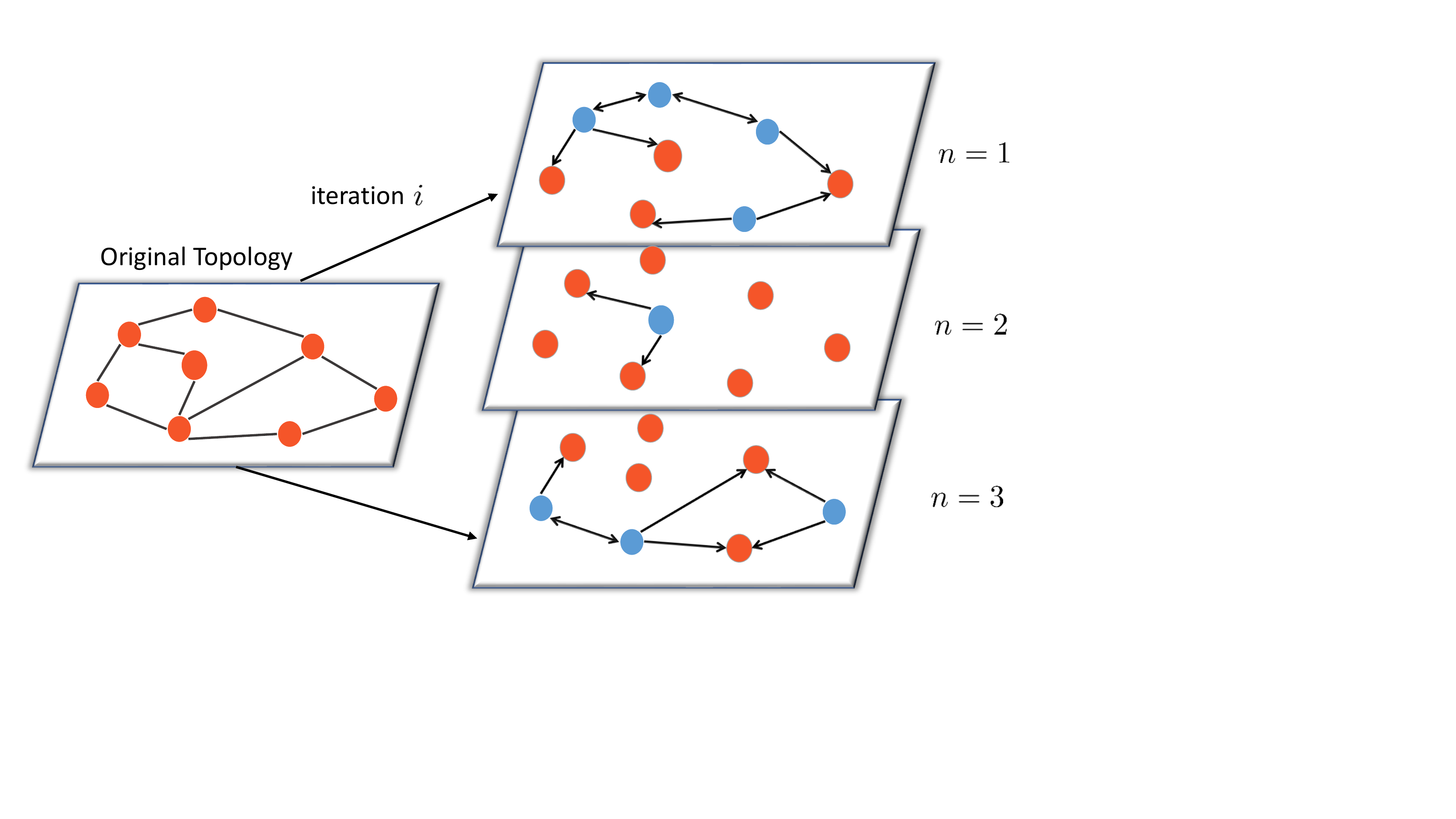}\vspace{-3mm}
	\caption{{\color{black}
			An example involving a network with $K=8$ and $N=3$. The sharing of information over the original network in a coordinate-wise manner can be viewed as sharing full information over a multi-layered topology.  The blue nodes at layer $n$ represent the agents that have activated entries at iteration $i$.
	}
}\label{fig.toplogy}
\end{figure}

In this way, agents now cannot share the same entries of their observation vectors. However, they will generally exist smaller groups of agents that end up selecting the same index (since indexes are chosen at random). We can represent this possibility by examining replicas of the network topology, as illustrated by Fig.~\ref{fig.toplogy}. In each layer, we highlight in blue the agents that selected the same index. For example, all four blue agents in the top layer selected the entry of index $n=1$; i.e., for these agents, $\n_i^k=1$. Only one agent in the second layer selected index $n=2$ and three agents in the bottom layer selected the index $n=3$.


\noindent Motivated by the discussion that led to Algorithm 3, we can similarly start from the following recursions: 
\eq{\everymath={\displaystyle}
	\hspace{-3mm}\begin{cases}
		\w_{k,i}(\n^k_i)&\hspace{-2mm}\!=  \hspace{-1mm}\sum_{\ell\in\cN_k, \n_{i}^\ell =\n^k_i}\hspace{-2mm}a_{\ell k}\Big(\w_{\ell,i-1}(\n_{i}^\ell) + r_{\ell,i}(\n_{i}^\ell) - \v_{\ell,i-1}(\n_{i}^\ell)\Big)\\
		\w_{k,i}(n) &\hspace{-2mm}= \w_{k,i-1}(n)\\
		&\hspace{5mm} +\hspace{-1mm}\!\!\! \sum_{\ell\in\cN_k, \n_{i}^\ell =n}\!\!\!\hspace{-2mm}a_{\ell k}\Big(\w_{\ell,i-1}(n) + r_{\ell,i}(n) - \v_{\ell,i-1}(n)\Big),\\
		&\hspace{4.5cm}{\rm if\ } n\neq\n_i^k \\
		\v_{k,i}(n) &\hspace{-2mm}= \begin{cases}
			\r_{k,i}(n),  &{\rm if\ } n=\n_i^k \\
			\v_{k,i-1}(n),  &{\rm if\ }n \neq \n_i^k 
		\end{cases}
	\end{cases} \raisetag{7mm}\label{jiong23ds}
}
where the summation $\sum_{\ell\in\cN_k, \n_{i}^\ell =n}$ refers to adding  over the neighbor agents $\ell$ whose selected random index $\n_i^\ell$ is equal to $n$. In this implementation, agents that select the same index within the neighborhood of agent $k$ are processed together in a manner similar to Algorithm 3. 
However, there is one important difficulty with this implementation, which does not work correctly. This is because  
%
%
\eq{
	\sum_{\ell\in\cN_k, \n_{i}^\ell =n}a_{\ell k}\neq \sum_{\ell\in\cN_k }a_{\ell k} = 1
}
In other words, the  ``effective'' combination matrix for any of the layers (on the right side of Fig. \ref{fig.toplogy}) is not necessarily doubly-stochastic anymore. Even worse, the topology from one layer to another and from one iteration to another keeps changing due the random selections at each agent. These facts bias the operation of the algorithm and prevent the agents from reaching consensus. We need to account for these difficulties. 

\subsection{Push-sum correction}
We shall exploit some properties from the push-sum construction.  Basically, recall that the  original push-sum algorithm deals with the problem of seeking the mean $\bar{r}$ of static signals $\{r_k\}$. One appealing property of the push-sum algorithm is that it can be applied to time-varying row stochastic matrices, i.e., to graphs where outgoing scaling factors add up to one, say,
\eq{
	\sum_{k=1}^K \a^{(i)}_{\ell k}=1,\;\;\;\;\forall \ell, i
}
where the superscript $i$ is added to indicate time-variation. This condition only requires the outgoing weights $\a_{\ell k}$ (from agent $\ell$ to agent $k$) to sum up to one; it does not require the incoming weights into agent $k$ to add up to one.  Moreover, it is common to assume that the topology  satisfies the following condition. 
\begin{assumption}[\sc Time-varying Topology Assumption\cite{benezit2010weighted} ]\label{assump.topology}
	The sequence $\A^{(i)}=[\a_{\ell k}^{(i)}]$ is a stationary and ergodic sequence of stochastic matrices with positive diagonal entries , and $\Ex \A^{(i)}$ is primitive.\qd
\end{assumption}
If we apply the classical consensus algorithm (\ref{consensus}) under this condition:
\eq{
	\w_{k, i} = \sum_{\ell \in \cN_k} \a^{(i)}_{\ell k} \w_{\ell,i-1},\;\;\;{\rm where\ } \w_{k,0} = r_k 
} 
then $w_{k, i}$ will not reach consensus \cite{xin2018linear}.
In order to reach consensus under this time-varing row stochastic topology, the push-sum algorithm construction introduces a vector variable $\p_{k,i}$ to help correct for bias. The algorithm starts from $\w_{k, 0} = r_k$ and $\p_{0,k} = \one$ ($\one$ is the vector with all entries equal one):\vspace{-1mm}
\eq{
	\left\{
	\begin{aligned}
	\w_{k,i} =& 	\sum_{\ell\in\cN_k} \a^{(i)}_{\ell k}\w_{\ell,i-1}\\
	\p_{k,i} =& 	\sum_{\ell\in\cN_k} \a^{(i)}_{\ell k}\p_{\ell,i-1}\\
	\x_{k,i} =& \w_{k,i} / \p_{k,i} \\[2mm]
	\end{aligned}\right.
}
where the last equality is used to mean that the individual entries of $\w_{k,i}$ are divided by the corresponding entry in $\p_{k,i}$; it refers to an element-wise division. It can be shown under Assumption \ref{assump.topology} that this algorithm leads to \cite{benezit2010weighted,nedic2015distributed}:
\eq{
	\lim_{i\to\infty} \x_{k,i} \stackrel{\rm a.s.}{=} \bar{r}
}	
Later in Sec. \ref{sec.push_sum}, we provide additional explanations that further clarify why this construction works correctly --- see the explanation leading to (\ref{push_sum.explain}).

{\color{black}
{\bf Remark}: Many other works on dynamic push-sum algorithms \cite{nedic2015distributed,notarnicola2018distributed,nedic2016stochastic} assume a B-strongly connected network (namely, that there exists a finite value $B$ such that the union of network topology in any consecutive $B$ iterations is strongly connected). This assumption does not fit into our case because we allow agents to select indexes independently of past selections and of other agents. Instead, we just the network to be strongly connected on average. \qd
}

\subsection{Dynamic diffusion with independent random updates}
We can exploit the push-sum construction in the dynamic diffusion scenario when random indexes are selected at each iteration. As we mentioned before, the implementation \eqref{jiong23ds} will not reach the desired consensus since the incoming weights $\{a_{\ell k}\}$ do not add up to one. However, we assumed the underlying matrix $A$ is doubly-stochastic, which implies that  
the outgoing weights still add up to one.  Hence, the push-sum construction can be utilized to solve the bias introduced by \eqref{jiong23ds}.
One important property to enforce  is that the entries in $p_{k,i}$ and $w_{k,i}$ should undergo similar updates. Doing so leads to Algorithm 4.

Comparing \eqref{dd.1}--\eqref{dd.2} with \eqref{sdd.1}--\eqref{sdd.2}, there are two main modifications.  One is that the updated index is allowed to vary at different locations. Another is that the output  is $\x_{k,i}$ instead of $\w_{k,i}$, i.e., the value after correction by $\p_{k,i}$. On the other hand, if we force $\n_{i}^k=\n_{i}^{k'}$ for all $k$ and $k'$, the Algorithm 4 will reduces to Algorithm 3 by noting that $\p_{k,i}=\one$ for any $i,k$. The proof of the convergence of Algorithm 3 is provided later in the Sec. \ref{sec.conv.alg3}.

\begin{table}
	\noindent\HRule\\
	\textbf{Algorithm 4 [Dynamic average diffusion algorithm with independent random updates]}\\[-2mm]
	\noindent\HRule\\
	{\bf Initialization:} \mbox{set $w_{k,0}=r_{k,0}; \, v_{k,0}=r_{k,0};\, p_{k,0}=\one.$} \\
	{\bf Repeat for $i=1,2,\ldots$ until convergence:}\vspace{-2mm}
	\begin{subequations}
		\begin{align}
		\n^k_i \sim&\, \cU[1,N] \hspace{6mm}\mbox{(locally uniform sampling)}\label{sdd.0}\\
		\w_{k,i} =&\, {\mathbb{S}}_{\n^k_i}^c\w_{k,i-1}\label{sdd.1}\\
		&\;\;+  \sum_{\ell \in \cN_k}a_{\ell k}\,\mathbb{S}_{\n^\ell_i}(\w_{\ell,i-1}+r_{\ell,i} - \v_{\ell,i-1})\nn\\
		\p_{k,i}=&\,{\mathbb{S}}_{\n^k_i}^c\p_{k,i-1} \sum_{\ell \in \cN_k}a_{\ell k}\,\mathbb{S}_{\n^\ell_i}\p_{\ell,i-1}\\
		\v_{k,i} =&\,  {\mathbb{S}}_{\n^k_i}^c\v_{k,i-1}\!+ \! \mathbb{S}_{\n^k_i}r_{k,i}\label{sdd.2}\\
		\x_{k,i} =& \w_{k,i} / \p_{k,i} \hspace{6mm}\mbox{(entry-wise division)}\label{sdd.3}
		\end{align}
	\end{subequations}
	\noindent\HRule\vspace{-2mm}
\end{table}

\subsection{Special case without push-sum correction}
There is one special case where $w_{k,i}$ from (\ref{jiong23ds}) can still  converge to the desired mean value without the push-sum correction. The special case is when \eq{
	\lim_{i\to\infty} \frac{1}{K}\sum_{k=1}^K r_{k,i} = 0
}
This scenario is quite common in the case of tracking the sum of gradients in empirical risk minimization problems. It can be verified that in this case it holds that  $\w_{k,i} \to 0$, i.e., with or without division by $\p_{k,i}$. {\color{black} To shed some intuition on this statement, assume that we have shown that the output $\x_{k,i}$ in Algorithm 4 has converged to the desired consensus value 0. 
 {\color{black} Then, we also know $\p_{k,i}$ is non-zero due to the non-zero initial value of $\p_{k,0}$ and the fact that  $A$ is primitive. Now from \eqref{sdd.3}, we conclude that 
	\eq{
		\w_{k,i}=\;&  \x_{k,i} \odot \p_{k,i} \nn\\
		=\;&  0 \p_{k,i} \nn\\
		=\;&0
	}
	where $\odot$ is the Hadamard product, i.e., entry-wise multiplication.  This confirms that without correction, all $\{\w_{k,i}\}$ will still converge to the desired value 0.
} 
The detailed proof of this statement is provided in the next section.

{\color{black} This also helps understand the recent push-pull type algorithm\cite{xin2018linear,pu2018push}, which uses a pull-network for consensus $\{w_{k}\}$ and a push-network for aggregating the gradient over agents. In that case, these works are interested in tracking the average of $\{\nabla J_k(w_k)\}$, where $J_k$ is some cost function associated with agent $k$. If the algorithm converges to the global optimal point of the aggregated cost function $\sum_{k=1}^K J_k(w)$, then we know that  he average of $\{\nabla J_k(w_k)\}$ should be zero and this conclusion is consistent with the situation just discussed. 
}

\section{CONVERGENCE ANALYSIS}
In this section, we establish the convergence of Algorithms 3 and 4 for both case of synchronous and independent random entry updates. 
\subsection{Convergence of Algorithm 3}\label{sec.conv.alg2}
First, we verify that recursions  \eqref{dd.1}--\eqref{dd.2} can reach consensus if the observation signals $r_{k,i}$ converge to $r_k$. Then we consider the case that the signals have a small perturbation. 

\begin{theorem}[\sc Mean-Square Convergence of Algorithm 3]\label{theorem.1}  Suppose the underlying topology $A$ satisfy the Assumption \ref{assumption-topology} and each signal $r_{k,i}$ converges to a limiting value $r_k$. It then holds that the algorithm converges in the mean-square-error sense, namely, 
	\eq{
		\lim_{i\to \infty} \Ex\|\w_{k,i} - \bar{r}_{i}\|^2 = 0,\;\;\; \forall k
	}
{\color{black}
The convergence rate depends on the rate at which the signals $\{r_{k,i}\}$ are varying. If the $\{r_{k,i}\}$ is a static signal, the convergence rate will be
\eq{
	\Ex\|\w_{k,i} - \bar{r}_{i}\|^2 \leq \alpha^i\Ex\|\w_{k,0} - \bar{r}_{0}\|^2, \;\;{\rm where}\; \alpha = 1 - \frac{1-\lambda}{N}	 \label{algorithm3.conv.rate}
}
and $\lambda$ is the second largest magnitude of the eigenvalue of $A$.
}
\end{theorem}
{\bf Proof}:
For a generic $n-$th entry of the weight/observation vectors, we collect their values into aggregate vectors as follows:
\eq{
	\sw_{i}(n) \define &\ba{c}\w_{1,i}(n)\\ \w_{2,i}(n)\\\vdots\\\w_{K,i}(n)\ea,\;\;\;
	{\sr}_{i}(n) \define \ba{c}{r}_{1,i}(n)\\ {r}_{2,i}(n)\\\vdots\\{r}_{K,i}(n)\ea\\
	&\;\;\;\;\;\sv_{i}(n) \define \ba{c}\v_{1,i}(n)\\ \v_{2,i}(n)\\\vdots\\\v_{K,i}(n)\ea
}
It is sufficient to establish convergence for one entry. Using the vector notation, we can verify that Algorithm 3 leads to: 
\eq{
	\sw_{i}(n) =\;& {\mathbb{S}}_{\n_i}^c(n) \sw_{i-1}(n)\nn\\
	& \;+{\mathbb{S}}_{\n_i}(n)  A \big(\sw_{i-1}(n)+\sr_i(n) -\sv_{i-1}(n)\big)\label{gjiosd}\\
	\sv_{i}(n) =\;& {\mathbb{S}}_{\n_i}^c(n)\sv_{i-1}(n) +{\mathbb{S}}_{\n_i}(n)\sr_i(n) \label{9h9h89ds}
}
where ${\mathbb{S}}_{\n_i}(n)$ is is a scalar and denotes the  $n-$th diagonal element of $\mathbb{S}$, which is either  $0$ or $1$ as defined in \eqref{fvowain}.
If we denote the average value vector by
\eq{
	\bar{\sw}_{i}(n) \define \frac{1}{K}\one\one\tran \sw_{i}(n)
}
we have
\eq{
	\bar{\sw}_{i}(n) =\;& {\mathbb{S}}_{\n_i}^c(n)\bar{\sw}_{i-1}(n) \label{giowd}\\
	&\!+{\mathbb{S}}_{\n_i}(n)\frac{1}{K}\one\one\tran\big(\sw_{i-1}(n)+\sr_i(n) -\sv_{i-1}(n)\big)\nn
}
where we used the property $\one\tran A = \one\tran$. 
Moreover, since the integer value of $\n_i$ is selected uniformly from the interval $[1,N]$ at iteration $i$, we have:
\eq{
	\Ex{\mathbb{S}}_{\n_i}^c(n)  &= \frac{N-1}{N}\\
	\Ex{\mathbb{S}}_{\n_i}(n)  &= \frac{1}{N}\\
	{\mathbb{S}}_{\n_i}(n){\mathbb{S}}_{\n_i}^c(n) &= 0
}
Because the following arguments focus on a single entry of index $n$, we shall drop $n$ for simplicity of notation. we can omit $(n)$.
Subtracting \eqref{gjiosd} from \eqref{giowd} and computing the expectation of the squared norm gives 
\eq{
	&\hspace{-6mm}\Ex\|{\boldsymbol{\sw}}_{i} - \bar{\sw}_{i}\|^2\nn\\
	 = &
	\frac{N-1}{N}\Ex\|\sw_{i-1}  - \bar{\sw}_{i-1} \|^2\nn\\
	&\;+	\frac{1}{N}\Ex\left\|(A-\frac{1}{K}\one\one\tran)\big(\sw_{i-1} +\sr_i  -\sv_{i-1} \big)\right\|^2\nn\\
	= &
	\frac{N-1}{N}\Ex\|\sw_{i-1}  - \bar{\sw}_{i-1} \|^2\nn\\
	&\;+	\frac{1}{N}\Ex\left\|\!\left(A\!-\!\frac{1}{K}\one\one\tran\right)\!\!\big(\!\sw_{i-1} \!-\!\bar{\sw}_{i-1} \!+\!\sr_i  -\sv_{i-1} \!\big)\!\right\|^2 \label{2gnio.sdwef}
}
where the last inequality exploits the fact  that 
\eq{
	\left(A-\frac{1}{N}\one\one\tran\right) \bar{\sw}_{i} = 0
}
Notice that under {\color{black} Assumption} \ref{assumption-topology}, we can show that matrix $A$ is primitive\cite{horn1990matrix, sayed2014adaptation} and, therefore, has one and only one  eigenvalue at one with its corresponding eigenvector equal to $\one$. 
Furthermore, the second largest magnitude of the eigenvalue of $A$, denoted by $\lambda$, satisfies\cite{pillai2005perron}:
\eq{
	0\leq\lambda<1
}When $\lambda=0$, which implies full-connectivity, we can end the proof quickly since \eqref{2gnio.sdwef} becomes:
\eq{
	\Ex\|{\boldsymbol{\sw}}_{i} - \bar{\sw}_{i}\|^2\leq \frac{N-1}{N}\Ex\|\sw_{i-1}  - \bar{\sw}_{i-1} \|^2
}
Hence, in the following argument, we exclude the trivial case $\lambda=0$.
We  continue with \eqref{2gnio.sdwef}  to get:
\eq{
	&\hspace{-2mm}\Ex\|{\boldsymbol{\sw}}_{i} - \bar{\sw}_{i}\|^2\nn\\
	\leq  &
	\frac{N-1}{N}\Ex\|\sw_{i-1}  - \bar{\sw}_{i-1} \|^2\nn\\
	&\;\;+	\frac{\lambda^2}{N}\Ex\left\|\sw_{i-1} - \bar{\sw}_{i-1} +\sr_i  -\sv_{i-1} \right\|^2\nn\\
	= &
	\frac{N-1}{N}\Ex\|\sw_{i-1}  - \bar{\sw}_{i-1} \|^2\nn\\
	&\;\;+	\frac{\lambda^2}{N}\Ex\left\|\frac{\lambda}{\lambda}(\sw_{i-1} - \bar{\sw}_{i-1}) +\frac{1-\lambda}{1-\lambda}(\sr_i  -\sv_{i-1}) \right\|^2\nn\\
	\leq& \frac{N-1}{N}\Ex\|\sw_{i-1}  - \bar{\sw}_{i-1} \|^2\nn\\
	&\;\;+	\frac{\lambda}{N}\Ex\left\|\sw_{i-1} - \bar{\sw}_{i-1} \right\|^2\!+\!\frac{\lambda^2}{N(1-\lambda)}\Ex\|\sr_i  -\sv_{i-1} \|^2\nn\\
	=& \Big(1-\frac{1-\lambda}{N}\Big)\Ex\|\sw_{i-1}  - \bar{\sw}_{i-1} \|^2+\frac{\lambda^2}{N(1-\lambda)}\Ex\|\sr_i  -\sv_{i-1} \|^2 \label{13gcxx}
}
where the second inequality is due to Jensen's inequality.
{\color{black}
	If the signal is static, we know $\Ex\|\sr_i  -\sv_{i-1} \|^2 = 0$ because $\sv_{i-1}$ is just the history record of signals $\sr_i$. Therefore, we can easily establish the convergence rate for static signal is $\alpha=1-\frac{1-\lambda}{N}$.
}

Similarly, we execute the same procedure on \eqref{9h9h89ds}:
\eq{
	&\hspace{-3mm}\Ex\|\sv_{i}  - \sr_{i+1} \|^2 \nn\\
	=\;&\Ex\big\|{\mathbb{S}}_{\n_i}^c(n)\sv_{i-1}   +{\mathbb{S}}_{\n_i}(n)\sr_i  - \sr_{i+1} \big\|^2\nn\\
	=\;&\Ex\|{\mathbb{S}}_{\n_i}^c(n)(\sv_{i-1}  - \sr_i )  + \sr_i  - \sr_{i+1} \|^2\nn\\
	=\;& \Ex\left\|\frac{t}{t}{\mathbb{S}}_{\n_i}^c(n)(\sv_{i-1}  - \sr_i )  + \frac{1-t}{1-t}\sr_i  - \sr_{i+1} \right\|^2\nn\\
	\leq\;& \frac{1}{t}\Ex\|{\mathbb{S}}_{\n_i}^c(n)\|^2\Ex\|\sv_{i-1}  - \sr_{i} \|^2 + \frac{1}{1-t}\|\sr_{i} - \sr_{i+1} \|^2\nn\\
	=\;&\frac{N-1}{N}\frac{1}{t}\Ex\|\sv_{i-1}  - \sr_{i} \|^2 + \frac{1}{1-t}\|\sr_{i} - \sr_{i+1} \|^2\nn\\
	=\;& \frac{N-1/2}{N}\Ex\|\sv_{i-1}  \!-\! \sr_{i} \|^2 \!+\! (2N-1)\|\sr_{i} \!-\! \sr_{i+1} \|^2 \label{23gdx}
}
where the inequality rely on  Jensen's inequality and we choose $t = \frac{N-1}{N-1/2}$ in last equality.
If $\sr_{i} $ converges, it means that
\eq{
	\|\sr_{i} - \sr_{i+1} \|^2 \to 0
}
so that due to \eqref{23gdx}, we conclude 
\eq{
	\Ex\|\sv_{i}  - \sr_{i+1} \|^2 \to 0
}
Combining with \eqref{13gcxx}, we get 
\eq{
\Ex\|\sw_{i}  - \bar{\sw}_{i} \|^2 \to 0 
}
Hence, we have proven that {\color{black} Algorithm 3} reaches the consensus if the observation signals converge. Lastly, we show that the consensus value is actually the desired $\bar{r}_i $. Let $\bar{\w}_i=\one\tran \sw_i/K$ and $\bar{\v}_i=\one\tran \sv_i/K$. it follows from \eqref{gjiosd} and \eqref{9h9h89ds} that
\eq{
	\bar{\w}_{i}  =&\;  {\mathbb{S}}_{\n_i}^c\bar{\w}_{i-1}  + {\mathbb{S}}_{\n_i}\big(\bar{\w}_{i-1} +\bar{r}_i  -\bar{\v}_{i-1} \big)\label{23g.84t}\\
	\bar{\v}_{i}  =& \; {\mathbb{S}}_{\n_i}^c\bar{\v}_{i-1}  + {\mathbb{S}}_{\n_i}\bar{r}_i \label{bnesd.g3}
}
Subtracting \eqref{bnesd.g3} from \eqref{23g.84t}, we have 
\eq{
	\bar{\w}_{i}  - \bar{\v}_{i}  = &\; {\mathbb{S}}_{\n_i}^c\left(\bar{\w}_{i-1}  - \bar{\v}_{i-1} \right)+{\mathbb{S}}_{\n_i}\left(\bar{\w}_{i-1}  - \bar{\v}_{i-1} \right)\nn\\
	=&\; \bar{\w}_{i-1}  - \bar{\v}_{i-1}  \label{keep.mean}
}
so that 
\eq{
	\bar{\w}_{i}  - \bar{\v}_{i}  =\bar{\w}_{0}  - \bar{\v}_{0}  = 0 \label{23g89c}
}
and we conclude $\bar{\w}_{i} $ is always the same as $\bar{\v}_{i} $. Recall that $\v_{k,i}$ is a vector that stores the past state of $r_{k,i}$ and it is easy to see that $\bar{\v}_{i} \to\bar{r}_{i} $ if $r_{i} $ converges, which completes the proof. \qd

\begin{corollary} [\sc Small Perturbations]
	Suppose each entry in the signal $r_{k,i}$ satisfies after sufficient iterations $i_o$:
	\eq{
		\|r_{k,i}(n) - r_{k,i-1}(n)\|^2 \leq \epsilon/N,\;\;\;\forall i>i_o, k \label{23gd}
	} 
	This property implies that $\|r_{k,i} - r_{k,i-1}\|^2 \leq \epsilon$, where $\epsilon $ is a small positive value. It then holds that
	\eq{
		\limsup_{i\to\infty} \frac{1}{K}\sum_{k=1}^K\Ex\|\w_{k,i}(n) - \bar{r}_{i}(n)\|^2 \leq \frac{2\lambda^2(2N-1)}{(1-\lambda)^2}\epsilon
	} 
\end{corollary}
{\bf Proof}: Substituting \eqref{23gd} into \eqref{23gdx},  for sufficiently large $i$, we have:
\eq{
	\Ex\|\sv_{i}  - \sr_{i+1} \|^2 
	\leq& \frac{N-1/2}{N}\Ex\|\sv_{i-1}  \!-\! \sr_{i} \|^2 \!+\! \frac{2N-1}{N}K\epsilon
}We omit $(n)$ again.
Taking the limit over $i$, we get
\eq{
	\limsup_{i\to\infty}\Ex\|\sv_{i-1}  \!-\! \sr_{i} \|^2\leq 2(2N-1)K \epsilon
}
Similarly, from \eqref{13gcxx}, we have
\eq{
	\limsup_{i\to\infty}	\Ex\|\sw_{i}  \!-\! \bar{\sw}_{i} \|^2 \leq& \frac{\lambda^2}{(1-\lambda)^2}\limsup_{i\to\infty}\Ex\|\sv_{i-1}  \!-\! \sr_{i} \|^2\nn\\
	\leq&\frac{2\lambda^2(2N-1)}{(1-\lambda)^2} K\epsilon
}
\qd

\subsection{Time-varying push-sum algorithm}\label{sec.push_sum}
Before we continue with the convergence proofs, we provide some useful intuition for the push sum construction. First, we note that the push-sum algorithm can be written in the following vector form for the $n-$th entry of the weight vectors (where we continue to drop the index $n$): 
\begin{subequations}
	\begin{align}
		\sw_{i} =& \big[\A^{(i)}\big]\tran \sw_{i-1} \label{push_sum.1}\\
		\scp_{i} =& \big[\A^{(i)}\big]\tran\scp_{i-1} \label{push_sum.2}\\
		\sx_{i} =& {\sw_{i}}/{\scp_{i}} \label{push_sum.3}
	\end{align}
\end{subequations}
where the last division is element-wise and 
\eq{
	\scp_{i}(n) \define \ba{c}\p_{1,i}(n)\\ \p_{2,i}(n)\\\vdots\\ \p_{K,i}(n)\ea
}
Recall that the combination matrix $A$ is row stochastic, which is equivalent to $\big[A^{(i)}\big]\tran$ is column stochastic, i.e.,
\eq{
	\one\tran  \big[A^{(i)}\big]\tran = \one\tran \label{23g.cdg3}
}
and satisfies Assumption \ref{assump.topology}.
It is shown in \cite{nedic2015distributed,benezit2010weighted,tahbaz2008necessary}.  that, for sufficient large $i$, 
\eq{
	\Big(\prod_{l=1}^i \big[A^{(l)}\big]\tran\Big)\define \big[A^{(i)}\big]\tran\big[A^{(i-1)}\big]\tran\cdots\big[A^{(1)}\big]\tran  \to \phi^i_{1}\one\tran
} 
where the stochastic vector $\phi^i_0$ (whose entries add up to one) keeps  changing with time no matter how large $i$ is. 
Then, it is easy to see when $i$ is sufficiently large:
\eq{
	\sw_{i} \to& \;\one\tran \sw_0\phi^i_1\\
	\scp_{i} \to& \;\one\tran \one\phi^i_1\\
	\sx_{i} \to& \; \bar{\w}_0\one\label{push_sum.explain}
}
Although $\phi^i_0$ keeps changing with time, the push-sum algorithm can reach consensus.

Before ending this section, we introduce a lemma that will be used in the convergence proof of the next section. 
\begin{lemma}[\sc Weak Ergodicity] \label{lemma.ergodic} Suppose the sequence of stochastic matrices $\{\A^{(l)}\}$ satisfies Assumption \ref{assump.topology} and for any $l$ and $l'$
	\eq{
		\Ex \A^{(l)} = \Ex \A^{(l')}\define A_E	
	}
	Then, there exists a unit vector $\bphi_j^i$ such that for any time index $i>j$:
	\eq{
		\Ex\Bigg\|\prod_{l=j}^i \big[\A^{(l)}\big]\tran - \bphi_j^i\one \tran\Bigg\|_{\rm max} \leq C\gamma^{i-j} \label{prod.A}
	}
	where $\|\cdot\|_{\rm max}$ means the element-wise maximum, and $C$ and {\color{black} $\gamma<1$} are constants that depend on the graph structure. 
	This means when $i-j$ is sufficiently large, the matrix converges to a rank-1 matrix, whose rows are identical. \vspace{-2mm}
\end{lemma}
{\bf Proof}:
{\color{black} Lemma} \ref{lemma.ergodic} is slightly different from the prior literature \cite{nedic2015distributed} and we therefore provide a sketch of the proof.\footnote{In \cite{nedic2015distributed}, it requires the topology to be strongly connected during any long enough duration $B$. In our case, this condition does not necessarily hold.}. The main idea is similar to  \cite{benezit2010weighted}. First, the Dobrushin coefficient $\delta(\A)$ of the column stochastic matrix $\A$ is defined as:
\eq{
	\delta(\A) \define \frac{1}{2}\max_{k,k'} \sum_{k=1}^K|\a_{\ell k} - \a_{\ell k'}|
}
To lighten the notation, we let 
\eq{
	\Q^{(j, i)} \define \prod_{l=j}^i \big[\A^{(l)}\big]\tran
}
It is easy to verify that:
\eq{
	\Ex\left\|\Q^{(j, i)} - \bphi_j^i\one \tran\right\|_{\rm max} \leq \Ex \delta\left(\Q^{(j, i)}\right)
}
so that we can focus on $\Ex \delta\left(\Q^{(j, i)}\right)$ instead.
It is shown in \cite{bremaud2013markov} that the Dobrushin coefficient has two useful properties \eq{
	\delta(\A^{(j)} \A^{(i)}) \leq \delta(\A^{(j)}) \delta(\A^{(i)})
} and
\eq{
	\delta(\A^{(i)}) \leq 1 - \max_{\ell} \min_{k} \a^{(i)}_{\ell k} \leq 1
}
Since we assume $\Ex A^{(j)}$ is primitive in Assumption \ref{assump.topology}, there exists a constant $T$ such that 
\eq{
	\Ex \Q^{(i-T, i)} =&\Ex \prod_{j=i-T}^i \big[\A^{(j)}\big]\tran\nn\\
	=& \prod_{j=i-T}^{i}\left[\Ex \A^{(j)}\right]\tran \nn\\
	=& \left(A_E\tran\right)^T\succ 0 \label{g23ds}
} 
where the notation $X \succ 0 $ means every entry of matrix $X$ is  strictly larger than 0. The strictly positive property is correct due to the primitive property on $A_E$ \cite{horn1990matrix,pillai2005perron,sayed2014adaptation}.
Hence, assuming $i-j\geq T$, we obtain:
\eq{
	\Ex \delta\left(\Q^{(j, i)}\right) \leq& \;\Ex \delta\left(\Q^{(j, i-T)}\right)\delta\left(\Q^{(i-T, i)}\right)\nn\\
	=& \;\Ex \delta\left(\Q^{(j, i-T)}\right) \Ex\delta\left(\Q^{(i-T, i)}\right)
}
Due to \eqref{g23ds}, there is at least one realization where all elements in one column of $\Q^{(i-T, i)}$ are strictly larger than 0, i.e.
\eq{
	\Pr\left[\delta\left(\Q^{(i-T, i)}\right) < 1\right] > 0
}
Combining the fact that $\delta\left(\Q^{(i-T, i)}\right) \leq 1$, we conclude that 
\eq{
	\Ex\delta\left(\Q^{(i-T, i)}\right) \define \gamma^{T}< 1
}
The exponent $T$ is just used for the purpose of simplifying the constant later.
Thus, for the case $i-j \geq T$ and supposing the modular representation $i = j + cT + r$, where $c\geq0$ and $T> r \geq0$, we  conclude
\eq{
	\Ex \delta\left(\Q^{(j, i)}\right) \leq \gamma^{cT}\Ex \delta\left(\Q^{(j, j+r)}\right)\label{eq.100}
}
Lastly, for the case $i-j< T$, we have
\eq{
	\Ex\delta\left(\Q^{(i, j)}\right) \leq 1 \leq C\gamma^{i-j} \label{eq.101}
}
where we let $C = (1/\gamma)^T$. Substituting \eqref{eq.101} into \eqref{eq.100}, we have:
\eq{
	\Ex \delta\left(\Q^{(j, i)}\right) \leq  \gamma^{cT} \cdot C\gamma^{r}  = C\gamma^{i-j}
}
\hfill\qd
\subsection{Convergence of Algorithm 4} \label{sec.conv.alg3}
\begin{theorem}[\sc Convergence of Algorithm 4]\label{theom.conv.ind} Suppose the underlying topology $A$ satisfy the Assumption \ref{assumption-topology} and each signal $r_{k,i}$ converges to $r_k$, then Algorithm 4 converges in the mean-square-error sense meaning that
	\eq{
		\lim_{i\to \infty} \Ex\|\x_{k,i} - \bar{r}_{i}\|_{\infty} = 0,\;\;\; \forall k
	}
{\color{black}
Similar to {\color{black} Theorem \ref{theorem.1}}, the convergence depends on the dynamic signal as well. If the signal is static, we have the linear convergence rate:
\eq{
	\Ex\|\x_{k,i} - \bar{r}_{i}\|_{\infty} \leq O(\gamma^{i})
}
where $\gamma$ is defined in \eqref{2g23,sd}.
}
\end{theorem}
{\bf Proof}:
Again, it is sufficient to focus on one entry/coordinate of the recursions. We have 
\eq{
	\sw_{i}(n) =& {\mathbb{K}}_i^{c} (n)\sw_{i-1}(n)\nn\\
	&\;\;+ A\tran\mathbb{K}_i(n)\big(\sw_{i-1}(n)+\sr_i(n) -\sv_{i-1}(n)\big)\nn\\
	=& \big[\A^{(i)}\big]\tran\sw_{i-1}(n) + A\tran\mathbb{K}_i(n)\big(\sr_i(n) -\sv_{i-1}(n)\big) \label{g34igvsd}
}
where
\eq{
	\mathbb{K}_i(n)  \define& \ba{cccc}
	\mathbb{I}[\n^1_i=n]&&&\\
	&\mathbb{I}[\n^2_i=n]&&\\
	&&\ddots&\\
	&&&\mathbb{I}[\n^K_i=n]
	\ea
	\\ 
	{\mathbb{K}}_i^{c} (n)\define& I_K - \mathbb{K}_i(n)\\
	[\A^{(i)}]\tran\define& {\mathbb{K}}_i^{(n)} + A\tran \mathbb{K}_i(n)  \label{define.A}
}
It is not hard to verify that $[\A^{(i)}]\tran$ is a time-varying column stochastic matrix as in \eqref{23g.cdg3}, and
\eq{
	\Ex [\A^{(i)}]\tran = \frac{N-1}{N} I + \frac{1}{N} A\tran
}
Obviously, $A^{(i)}$ satisfies Assumption \ref{assump.topology}. Similarly, we have
\eq{
	\scp_{i}(n)=& {\mathbb{K}}_i^{c} (n)\scp_{i-1}(n)+ A\tran {\mathbb{K}}_i(n)\scp_{i-1}(n)\nn\\
	=& \big[\A^{(i)}\big]\tran\scp_{i-1}(n)\\
	\sv_{i}(n) =& {\mathbb{K}}_i^{c} (n)\sv_{i-1}(n) +{\mathbb{K}}_i(n)\sr_i(n)\nn\\
	=& \sv_{i-1}(n) +{\mathbb{K}}_i(n)(\sr_i(n) - \sv_{i-1}(n))\label{gi2}
}
Substituting (\ref{gi2}) into (\ref{g34igvsd}), we have
\eq{
	\sw_{i}(n) =& \big[\A^{(i)}\big]\tran\sw_{i-1}(n) + A\tran (\sv_{i}(n) - \sv_{i-1}(n))\label{recursio1}
}
Next, we establish the same result as \eqref{keep.mean} by computing  the sum of \eqref{recursio1}: 
\eq{
	\one\tran\sw_{i}(n) =& \one\tran\sw_{i-1}(n) + \one\tran(\sv_{i}(n) - \sv_{i-1}(n))\nn\\
	=& \one\tran\sw_{i-2}(n) + \one\tran(\sv_{i-1}(n) - \sv_{i-2}(n)) 
	\nn\\
	&\;\;\;+ \one\tran(\sv_{i}(n) - \sv_{i-1}(n))\nn\\
	=&\one\tran\sw_{0}(n) + \one\tran\sv_{i}(n) - \one\tran\sv_{0}(n)\nn\\
	=&  \one\tran\sv_{i}(n)
}
where the last equality is because in the algorithm, we use $w_{k,0} = v_{k,0}$ so that $\one\tran\sw_{0}(n) - \one\tran\sv_{0}(n) = 0$.
Similarly, we have
\eq{
	\one\tran \scp_i(n) =& \one\tran \big[\A^{(i)}\big]\tran\scp_{i-1}(n)\nn\\
	=& \one\tran\scp_{i-1}(n)\nn\\
	=& \one\tran\scp_{0}(n)\nn\\
	=& N \label{23.h24}
}
Let
\eq{
	\alpha_{i}(n) \define \frac{\one\tran\sw_{i}(n)}{N} = \frac{\one\tran\sv_{i}(n)}{N} \label{12,g23gswd}
}
Starting below, we will ignore $(n)$ for simplicity:
 \eq{
 	&\hspace{-5mm}\sw_{i} - \alpha_{i}\scp_{i}\nn\\
 	=& \big[\A^{(i)}\big]\tran (\sw_{i-1} - \alpha_{i}\scp_{i-1}) + A\tran(\sv_{i} -\sv_{i-1})\nn\\
 	=& \big[\A^{(i)}\big]\tran (\sw_{i-1} - \alpha_{i-1}\scp_{i-1}) \nn\\
 	&\;\;+ \big[\A^{(i)}\big]\tran\alpha_{i-1}\scp_{i} - \big[\A^{(i)}\big]\tran\alpha_{i}\scp_{i} + A\tran(\sv_{i} -\sv_{i-1}) \label{23fg}
 }
Let\vspace{-1mm}
\eq{
	\z_i \define (\alpha_{i-1} - \alpha_{i})\big[\A^{(i)}\big]\tran\scp_{i}+ A\tran(\sv_{i} -\sv_{i-1}) \label{define.z}
}
then recursion \eqref{23fg} becomes:
\eq{
	\sw_{i} - \alpha_{i}\scp_{i} =  \big[\A^{(i)}\big]\tran (\sw_{i-1} - \alpha_{i-1}\scp_{i-1}) + \z_i \label{g2jtred}
}
It is easy to verify that 
\eq{
	\one\tran \z_i =& (\alpha_{i-1} - \alpha_{i})\one\tran\big[\A^{(i)}\big]\tran\scp_{i}+ \one\tran A\tran(\sv_{i} -\sv_{i-1})\nn\\
	\stackrel{\eqref{23g.cdg3}}{=}&(\alpha_{i-1} - \alpha_{i})\one\tran\scp_{i}+ \one\tran (\sv_{i} -\sv_{i-1})\nn\\
	\stackrel{\eqref{23.h24}}{=}&(\alpha_{i-1} - \alpha_{i})N+(\alpha_{i} - \alpha_{i-1})N\nn\\
	=&0\label{29,mg23}
} To give some intuition, if $\sv_i$ converges, then we have:
\eq{
	\alpha_{i-1} - \alpha_{i}\to0,\;\;\;\sv_{i-1} - \sv_{i}\to 0,\;\;\; \z_i \to 0 \label{epsilon.convg}
}
Recall that $\sv_i$ is the history record of signal $r_i$, which implies that if the signals gradually converge, then iteration \eqref{g2jtred} will eventually be equal to the consensus algorithm.
Now, expanding \eqref{g2jtred} with respect to $i$, we get\vspace{-1mm}
\eq{
	&\hspace{-4mm}\sw_{i} - \alpha_{i}\scp_{i} \nn\\
	=& \left(\prod_{l=1}^i \big[\A^{(i)}\big]\tran\right)(\sw_{0} - \alpha_{0}\scp_{0}) + \sum_{j=1}^{i}\left(\prod_{l=j+1}^i \big[\A^{(i)}\big]\tran\right)\z_j \label{g23gsd}
}
Recalling the definition in Eq. \eqref{12,g23gswd} and the property in Eq. \eqref{29,mg23}, we have
\eq{
\phi_0^i\one\tran(\sw_{0} - \alpha_{0}\scp_{0})&=0,\;\;\;\forall i\\
 \phi_j^i\one\tran \z_j&=0\;\;\;\;\;\forall i, j
}
so that expression \eqref{g23gsd} is equivalent to \vspace{-1mm}
\eq{
	\sw_{i} - \alpha_{i}\scp_{i} =& \left(\prod_{l=1}^i \big[\A^{(i)}\big]\tran - \bphi_0^i\one\tran\right)(\sw_{0} - \alpha_{0}\scp_{0})\nn\\
	&\;\; + \sum_{j=1}^{i}\left(\prod_{l=j+1}^i \big[\A^{(i)}\big]\tran - \bphi_j^i\one\tran \right)\z_j \label{23gsD}
}

\noindent We further introduce the notation\eq{
	[\B_{j}^i]\tran \define& \prod_{l=j}^i \big[\A^{(i)}\big]\tran - \bphi_j^i\one \tran\\
	[\B_{j}^i]\tran_k \define& \mbox{the $k-$th row of } \prod_{l=j}^i \big[\A^{(i)}\big]\tran - \bphi_j^i\one \tran
}
It is straightforward to show that the sequence of matrices $\{A^{(l)}\}$ defined in \eqref{define.A} satisfies  all the assumptions in Lemma \ref{lemma.ergodic}. Hence, for some constant $C$ we have
\eq{
	\Ex\Big\|[\B_{j}^i]\tran\Big\|_{\rm max} \leq&\; C\gamma^{i-j}\\
	\Ex\Big\|[\B_{j}^i]_k\Big\|_{\infty} \leq & \;C\gamma^{i-j} \label{9.321.23g} ,\;\;\;\;\forall k
}
Taking the infinity norm of \eqref{23gsD} and expectation, we have
\eq{
	&\Ex\|\sw_{i} - \alpha_{i}\scp_{i}\|_{\infty}\nn\\
	& \leq \Ex\left\|[\B_{j}^i]\tran (\sw_{0} - \alpha_{0}\scp_{0})\right\|_\infty + \Ex\left\|\sum_{j=1}^{i}[\B_{j}^i]\tran\z_j\right\|_\infty\nn\\
	& = \Ex \max_k \Big|[\B_{1}^i]\tran_{k}(\sw_{0} - \alpha_{0}\scp_{0}) \Big| + \Ex\max_k \Big|\sum_{j=1}^i [\B_{j}^i]\tran_{k}\z_j \Big| \nn\\
	& \stackrel{(a)}{\leq} \Ex\max_k \|[\B_{1}^i]_{k}\|_\infty \|\sw_{0} - \alpha_{0}\scp_{0}\|_1 + \Ex\max_k \sum_{j=1}^i\|[\B_{j}^i]_{k}\|_\infty \|\z_j\|_1 \nn\\
	&\stackrel{(\ref{9.321.23g})}{\leq}C\gamma^{i-1}\|\sw_{0} - \alpha_{0}\scp_{0}\|_1 + \sum_{j=1}^{i}C\gamma^{i-j}\|\z_j\|_1 \label{2389d.sg}
}
where step (a) exploits the Holder inequality\cite{Kolmogorov1991elements, boyd2004convex}:
\eq{
	|x\tran y| \leq \|x\|_\infty \|y\|_1
} 
Finally, supposing for any $\delta>0$, there exists an $N$ such that $\|\z_i\|_1\leq \delta,\;\;i>N$ due to \eqref{epsilon.convg}, we have 
\eq{
	&\|\sw_{i+1} - \alpha_{i+1}\scp_{i+1}\|_{\infty}\nn\\
	&\leq C \gamma^{i-1}\|\sw_{0} - \alpha_{0}\scp_{0}\|_1 + \sum_{j=1}^{N}C \gamma^{i-j}\|\z_j\|_1 + \sum_{j=N}^{i}\gamma^{i-j}\delta\nn\\
	&\leq C \gamma^{i-1}\|\sw_{0} - \alpha_{0}\scp_{0}\|_1 + \gamma^{i-N}\sum_{j=1}^{N}C \gamma^{N-j}\|\z_j\|_1 + \frac{1}{1-\gamma}\delta
}
Letting $i\to\infty$, we have
\eq{
	\lim_{i\to\infty}\|\sw_{i} - \alpha_{i}\scp_{i}\|_{\infty} \leq \frac{1}{1-\gamma}\delta \label{1241}
}
Since $\delta$ can be arbitrarily  close to 0, we conclude that:
\eq{
		\lim_{i\to\infty}\|\sw_{i}/\scp_{i} - \alpha_{i}\one\|_{\infty} = 0
}

Lastly, we show that if $\sum_{k=1}^K r_{k,i} \to 0$, then $\w_{k,i}\to 0$ for all $k$.
Using the triangle inequality and  \eqref{1241}, we obtain
\eq{
		\lim_{i\to\infty}\|\sw_{i}\|_{\infty}&\leq \lim_{i\to\infty}\|\sw_{i} - \alpha_{i}\scp_{i}\|_{\infty}
		+\lim_{i\to\infty}\|\alpha_{i} \scp_{i}\|_{\infty} \nn\\
		&\leq \frac{1}{1-\gamma}\delta +|\alpha_{i}|\lim_{i\to\infty}\|\scp_{i}\|_{\infty}
} 
Because $\alpha_{i}$ is the desired average value, we have $\alpha_{i}\to0$. Therefore, we conclude all  $\w_{k,i}$ across the agents will converge to zero.

{\color{black}
Similarly, we can obtain the convergence rate when the signal is static although the argument is not as concise as in the synchronized case.   First, recall $\sv_i$ is just the history record of signal $r_i$. It is not hard to see when the signal is static, the variable $\z_i$ defined in \eqref{define.z} will be zero. Then, \eqref{2389d.sg} is simplified to 
\eq{
	\Ex\|\sw_{i} - \alpha_{i}\scp_{i}\|_{\infty} \leq C\gamma^{i-1}\|\sw_{0} - \alpha_{0}\scp_{0}\|_1
}
where 
\eq{
	\gamma \define \sqrt[\leftroot{-1}\uproot{5}T]{\xi},\;\;\; {\rm where} \;\;\xi =\Ex \delta\left(\prod_{i=1}^T A^{(i)}\right)\label{2g23,sd}
} 
Recall that the Dobrushin coefficient of the column stochastic matrix $\A$ is defined as\cite{bremaud2013markov}:
\eq{
	\delta(\A) \define \frac{1}{2}\max_{k,k'} \sum_{k=1}^K|\a_{\ell k} - \a_{\ell k'}|
}
As we have shown in Lemma \ref{lemma.ergodic}, $\gamma$ is a positive number that is strictly smaller than zero. Lastly, notice that 
\eq{
	\Ex\|\sw_{i} - \alpha_{i}\scp_{i}\|_{\infty} =& \Ex\|(\sx_{i}- \alpha_{i}) \odot \scp_{i} \|_{\infty} \nn\\
	\geq& \left(\min \scp_{i} \right) \Ex\|\sx_{i}- \alpha_{i}\|_{\infty} \label{2389g/sd}
}
where $\min$ represents the smallest element of $\scp_{i}$. Recall that we know $\p_{k,i}$ is non-zero due to the non-zero initial value of $\p_{k,0}$ and the fact $A$ is primitive. Therefore, any element in $\min \scp_{i} $ is strictly positive. Combining all above results, we conclude that 
\eq{
		\Ex\|\sx_{i}- \alpha_{i}\|_{\infty} \leq O(\gamma^i)
}
Lastly, notice in static signal scenario $\alpha_{i}$ is always equal to the real average, it is equivalent to conclude that 
\eq{
	\Ex\|\x_{k,i} - \bar{r}_{i}\|_{\infty} \leq O(\gamma^{i})
}
Unfortunately, the convergence rate of {\color{black}Algorithm 4} does not have a closed form expression so that it is hard to compare with {\color{black} Algorithm 3}. However, in the next numerical simulation section, we will observe that the convergence rate of {\color{black}Algorithm 4} is still similar to the others.
\qd
}

\section{NUMERICAL SIMULATION} \label{sec.simulation}
{\color{black} The simulations in this section are all based on the same setting, unless otherwise stated.  We generate a network with $25$ agents, as shown in  Fig. \ref{fig.topology}.} The dimension $N$ is set to $N=100$ and each entry of $r_{k,i}$ is generated according to the following model:{\color{black}
\eq{
	r_{k,i}(n) = \a_k(n)\exp^{-\alpha i}\sin(\beta i) + \b_k(n) + \gamma i \label{9.123}
}
where $\a_k(n)$ and $\b_k(n)$ are zero-mean Gaussian distributed with variance 1. The parameters $\alpha$, $\beta$, and $\gamma$ are set to $\alpha=0.01$, $\beta=0.1$ and $\gamma=2.5e-4$. It is seen that, as the iteration index $i$ increases, the signals $r_{k,i}(n)$ converge to $\b_{k}(n)+\gamma i$. We generate 2000 samples according to model (\ref{9.123}) and at iteration $i=2000$, we replace $\gamma i$ by $-\gamma i$ in (\ref{9.123}) in order to change the dynamics of signals. This will enable us to observe the tracking mechanism by the dynamic-average diffusion strategy. The sampled signals are plotted in Fig.~\ref{signals.curve}.}

 \vspace{-2mm}
\begin{figure}[!htp]
	\centering
	\includegraphics[scale=0.38]{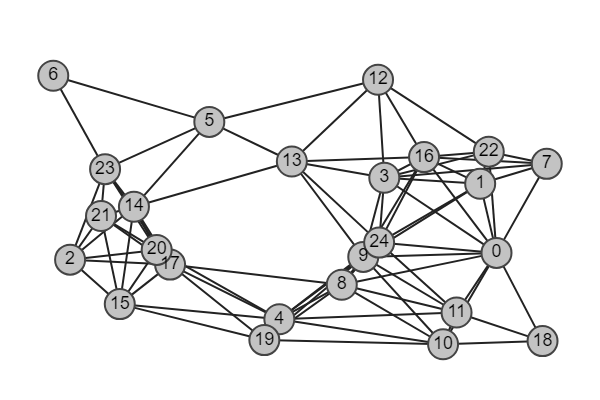}\vspace{-2mm}
	\caption{ Simulated network topology. \label{fig.topology}}
\end{figure}

\begin{figure}[!htp]
	\centering
	\includegraphics[scale=0.5]{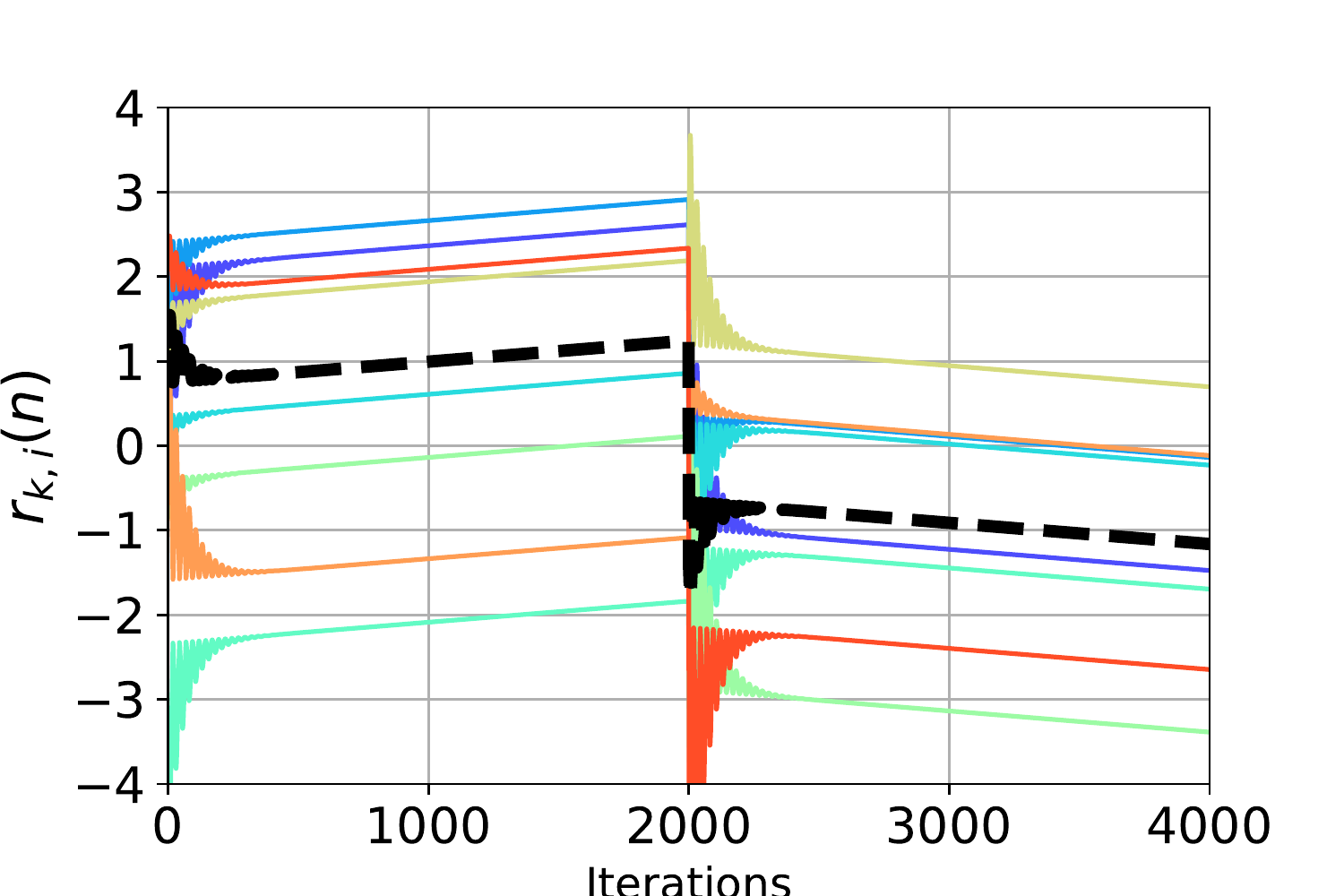}
	\caption{Each solid line represents the one dimension of observed signals for some of the agents (not all agents are shown in order to avoid overcrowding the figure),  and the dotted line marks the average of the signals from across all agents.\label{signals.curve}}
\end{figure}

{\color{black}
\subsection{Comparison of Dynamic Average Algorithms}
As was already mentioned in Section \ref{sec.other.algorithm}, we can derive many variations of dynamic average algorithms from different distributed gradient algorithms. We examine the tracking performance of those algorithms based on the 25 agents topology and the dynamic signals described by \eqref{9.123}. Figure \ref{fig.diff_algs} shows that the diffusion-based algorithm \eqref{dynamic.diffusion}, consensus-based algorithm \eqref{extra-based.alg}, EXTRA-based algorithm \eqref{consensus-based.alg} are quite similar and converge faster than the DIGing based algorithm. From the zoom-in sub-figure, we see that the diffusion based algorithm has some slight performance boost.

}

\begin{figure}[tp]
	\centering
	\includegraphics[scale=0.42]{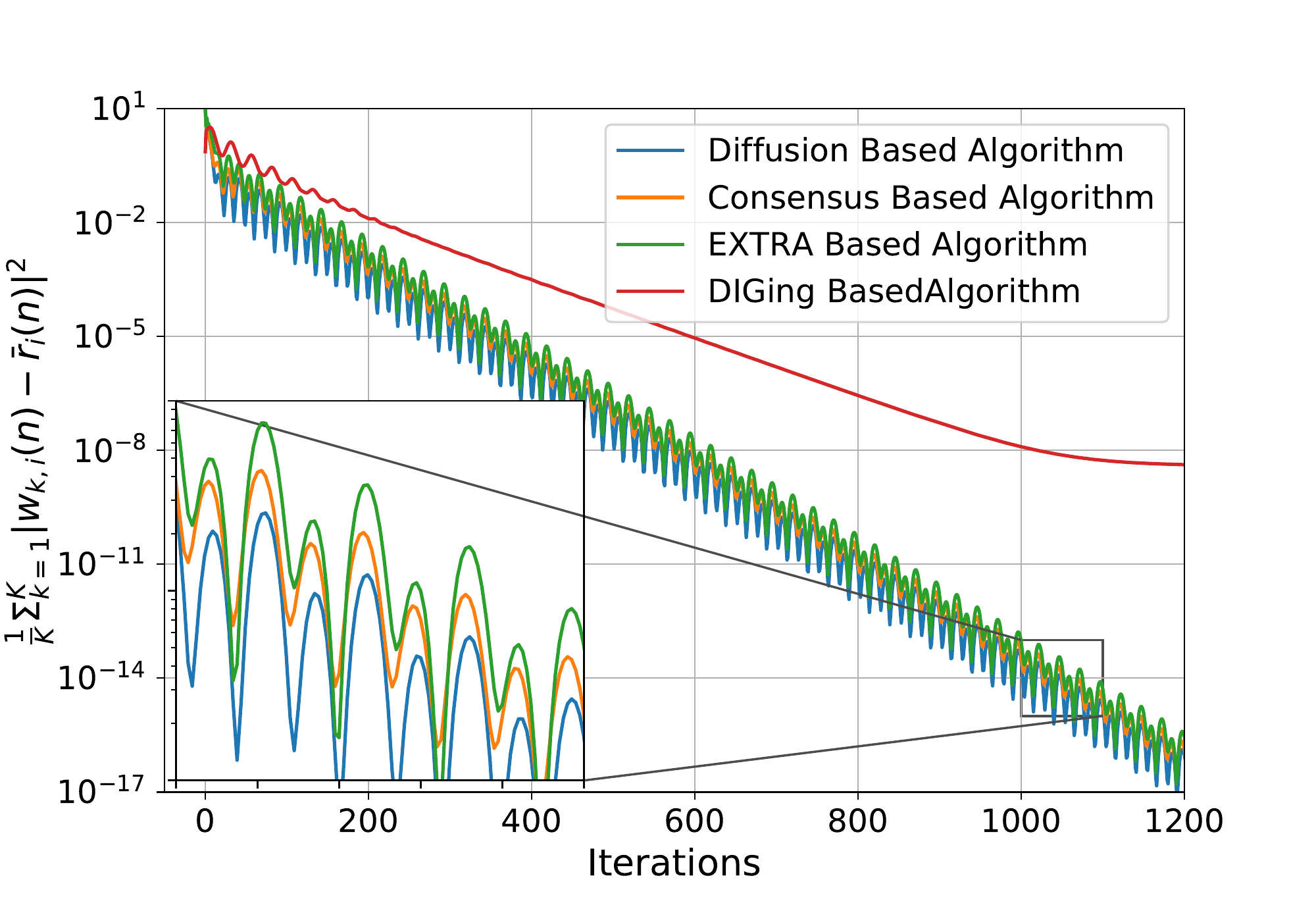}
	\caption{\color{black} Comparison with other related techniques. Diffusion based algorithm refers to \eqref{dynamic.diffusion}, consensus based algorithm refers to\eqref{extra-based.alg}, EXTRA based algorithm refers to\eqref{consensus-based.alg}, and DIGing based algorithm refers to \eqref{2378sdbsd9}. \label{fig.diff_algs}}
\end{figure}

\subsection{Full Update, Synchronous Random Update,  and Independent Random Update}
{\color{black} Next, we illustrate the results of Theorems \ref{theorem.1} and \ref{theom.conv.ind}  by means of numerical simulations. We will see that independently and randomly updating only one entry does not harm  performance in terms of convergence rate by communicated length.} We plot in Fig. \ref{conv.curve} the error measure  $\frac{1}{K}\sum_{k=1}^K\|\w_{k,i}(n) - \bar{r}_i(n)\|^2$ only for the first entry, i.e., for $n=1$ for illustration purposes. It is seen that this measure decreases, as expected, and that the network is able to track the new average value after the perturbation at $i=2000$. 

The figure shows three curves: one corresponding to synchronous updates where all agents select the same entry of the iterates to communication, one corresponding to asynchronous updates where different agents may select randomly different entries, and the original dynamic average consensus algorithms from\cite{freeman2006stability, zhu2010discrete}, i.e., the full-length versions. 

Observe that, at each iteration, each agent will send only one entry to his neighbors in our proposed algorithm while agents in full dynamic average algorithm will send $N$-length vectors. Hence, a fair comparison is based on the total communicated vector length\cite{richtarik2014iteration}. Although we can easily compute the convergence rate for static signals in {\color{black} Algorithm 3}, it is still informative. From \eqref{algorithm3.conv.rate} we know, after $N$-iterations, the convergence rate is 
\eq{
	\alpha^N = (1-\frac{1-\lambda}{N})^N \approx \lambda	
}
where the last approximation holds if $N$ is large enough, which is reasonable in our case. That $\lambda$-rate is the same rate for the full-length dynamic consensus algorithm. Therefore, we observe in Fig.~\ref{conv.curve} that the convergence rate for our proposed {\color{black}Algorithms 3, 4}, and the original dynamic consensus are similar. 

Further, we observe that the curve for synchronous updates has a typical stair-like shape while the one for independent updates does not. The stair-like shape is due the coordinate-wise updates, which imply that the error would stay constant until the coordinate is selected again. 

\begin{figure}[!htp]
	\centering
	\includegraphics[scale=0.44]{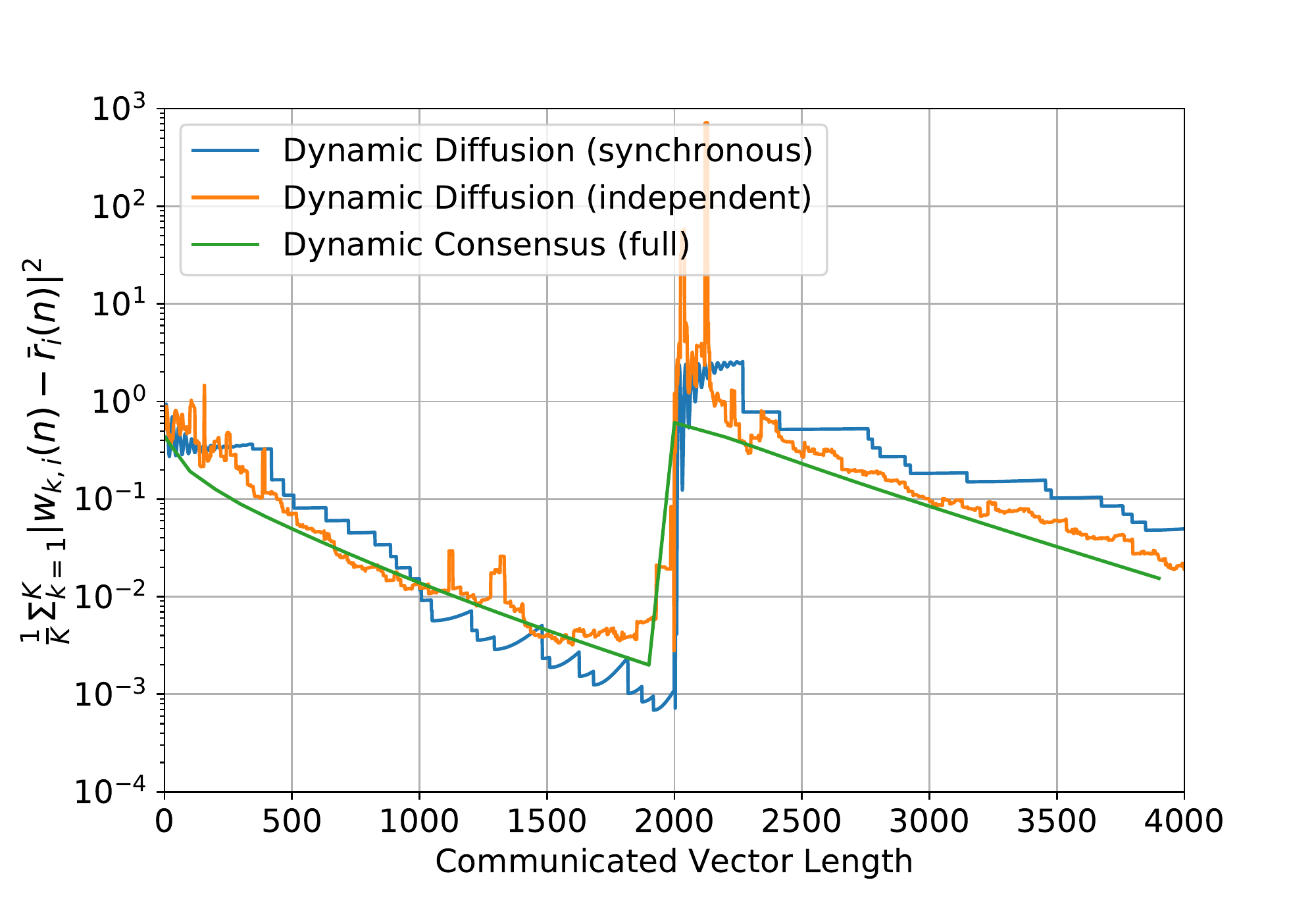}
	\caption{The convergence curve of dynamic average diffusion with both synchronous, i.e., Eqs.(\ref{dd.1}) -- (\ref{dd.3}), asynchronous updates, i.e., Eqs.\eqref{sdd.0} -- \eqref{sdd.3}, and dynamic consensus in \cite{freeman2006stability}. \label{conv.curve}}
\end{figure}


{\color{black}
\subsection{Effect of Topology on Performance}
We indicated before that the convergence rate of Algorithm 3 is dependent on the network topology $\lambda$ and the length of signal dimension $N$ ---- \eqref{algorithm3.conv.rate}. In a similar vein, it is possible to predict that the behavior of Algorithms 3 and 4 in the dynamic signal scenario should follow a similar pattern.

The simulations vary the number of agents, i.e., the number of nodes in the network, based on two topology styles: 
\begin{enumerate}
	\item Random geometric networks with fixed connection radius.  In this case, each agent is placed randomly in  a $[0, 1]\times[0, 1]$ square. As long as the distance between two agents is shorter than a certain threshold, these two agents are connected. Therefore, it is not hard to see that with more agents generated, the connectivity of the network gets better which, in general, means $\lambda$ is getting smaller, $\lambda_{\rm geo} = O(1/K)$ \cite[Sec. 5-2]{mao2018walkman}.
	
	\item Fixed cyclic networks. In this case, agent $k$ only connects with agent $k-1$ and agent $k+1$. When there are more agents, the network becomes more sparse and $\lambda$ gets closer to 1 quickly. Actually, we have analytic expression that $\lambda_{\rm ring}(K) = 1-O(K^{-2})$.
\end{enumerate}
These two types of networks are illustrated in Fig.~\ref{fig.different_topo}. We examine the performance of Algorithm 4  with different numbers of agents. As we have shown in Theorem \ref{theorem.1}, the convergence rate is related to the network's connectivity. The dynamic signal setting is same as the one stated before in  \eqref{9.123} except that we no longer change the dynamics of signals in the middle.

The simulation results are shown in Figs.~\ref{fig.err_rgg} and \ref{fig.err_cycle}. We can get several useful observations from them: 1. The algorithm converges much faster in a dense network than a sparse network.  2. The convergence rate has negative correlation with the second largest eigenvalue. However, the relation is not linear or inverse linear. 3. The algorithm converges at different speeds at different times, i.e., faster in the beginning and then slower.

\begin{figure}[!htp]
	\centering
	\includegraphics[scale=0.29]{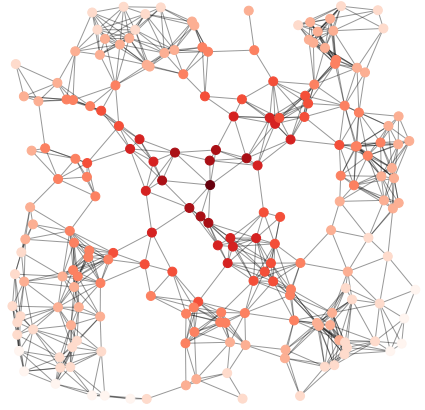}\hspace{-0mm}
	\includegraphics[scale=0.26]{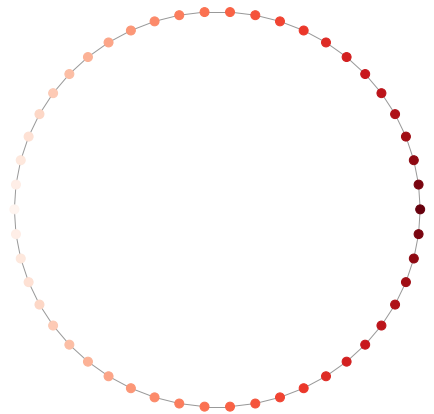}
	\vspace{-2mm}
	\caption{\color{black} Random geometric network (left) and cyclic network (right). \label{fig.different_topo}}
\end{figure}

\begin{figure}[!htp]
	\centering
	\includegraphics[scale=0.4]{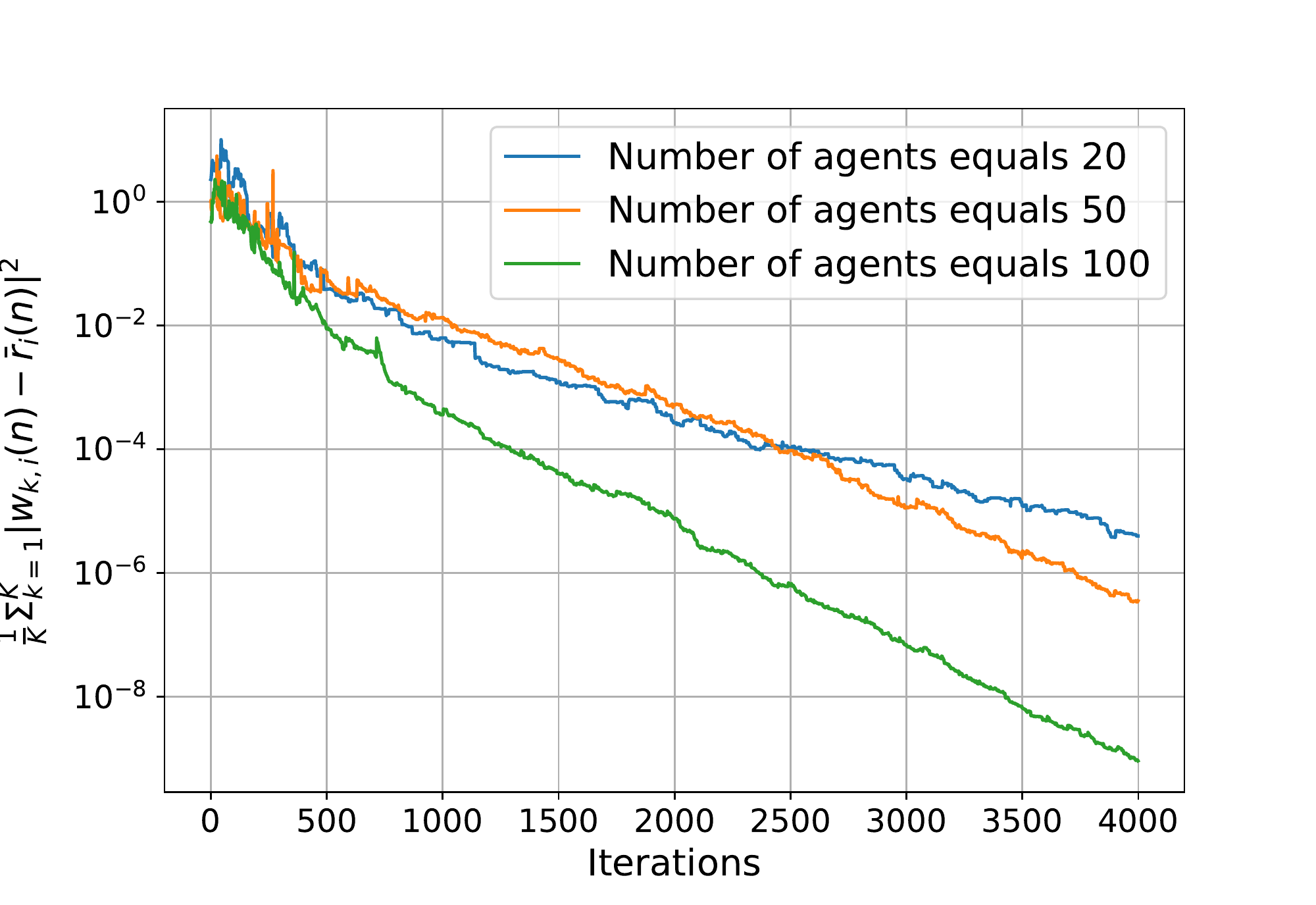}
	\vspace{-1mm}
	\caption{\color{black} Convergence of Algorithm 4 with different numbers of agents on  random geometric networks. The second largest eigenvalue is $0.89$ for 20 agents, $0.83$ for 50 agents, and $0.79$ for 100 agents.\label{fig.err_rgg}}
\end{figure}

\begin{figure}[!htp]
	\centering
	\includegraphics[scale=0.4]{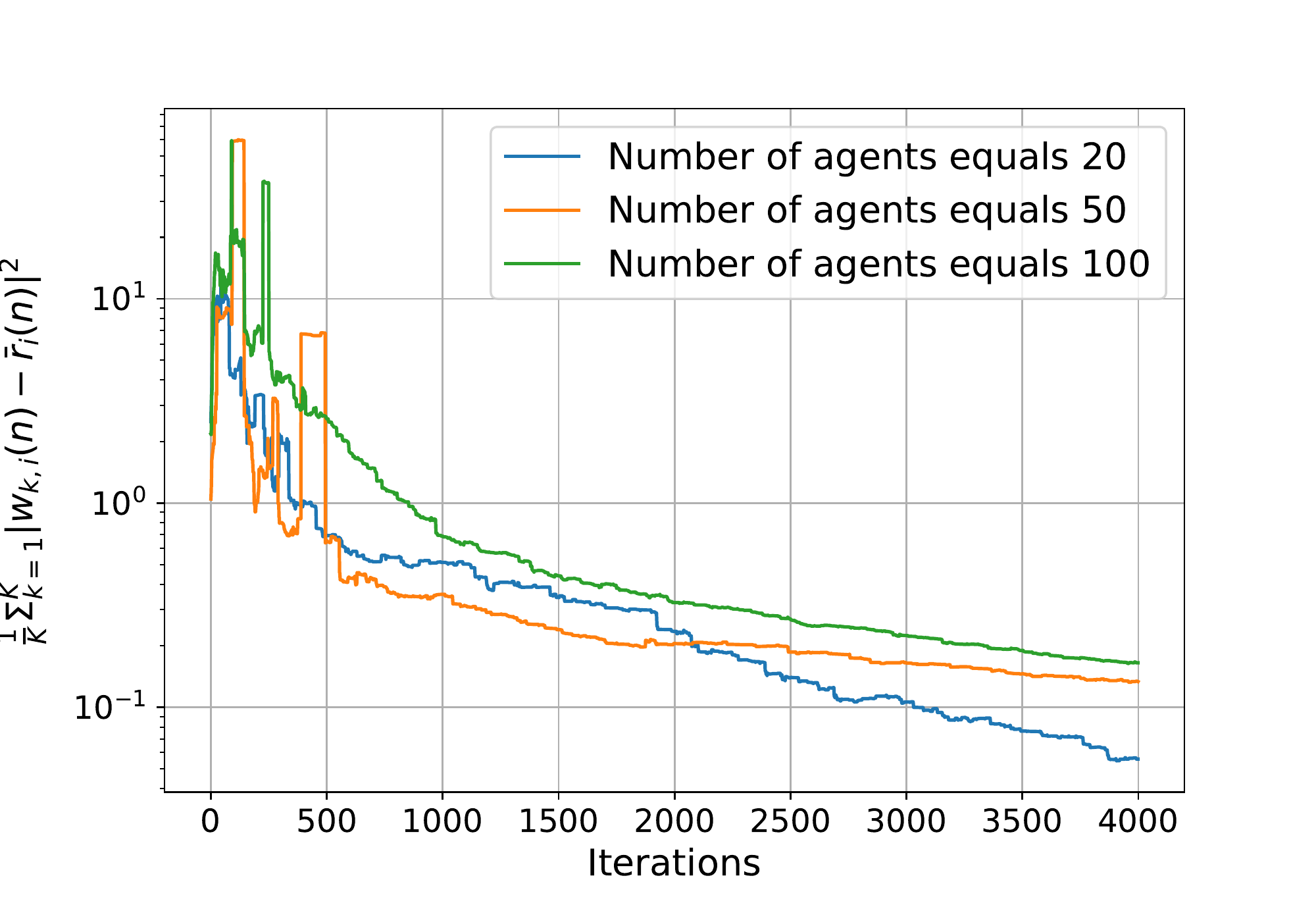}
	\vspace{-1mm}
	\caption{ \color{black} Convergence of Algorithm 4 with different numbers of agents over the cyclic networks. The second largest eigenvalue is $0.967$ for 20 agents, $0.995$ for 50 agents, and $0.998$ for 100 agents.\label{fig.err_cycle}}
\end{figure}

}

\section{Conclusion and Future Works} 
In summary, this works derives and analyzes an online learning strategy for tracking the average of time-varying distributed signals by relying on randomized coordinate-descent updates. We proposed two dynamic-average diffusion algorithms: in one case all agents select the same entry from the observations, and in the second case all agents may select different entries from their observations. Auxiliary variables and push-sum ideas are utilized to avoid bias and ensure convergence.

{\color{black}Future work involves applying the proposed techniques to scenarios dealing with distributed features or gradient boosting learning. While this work focused on quadratic costs, one can consider extensions to other cost functions as well. } 

\bibliographystyle{IEEEbib}
\bibliography{dynamic_diffusion}

\begin{thebibliography}{10}

\bibitem{sayed2014adaptation}
A.~H. Sayed,
\newblock ``Adaptation, learning, and optimization over networks,''
\newblock {\em Foundations and Trends in Machine Learning}, vol. 7, no. 4--5,
  pp. 311--801, 2014.

\bibitem{nedic2017achieving}
A.~Nedic, A.~Olshevsky, and W.~Shi,
\newblock ``Achieving geometric convergence for distributed optimization over
  time-varying graphs,''
\newblock {\em SIAM Journal on Optimization}, vol. 27, no. 4, pp. 2597--2633,
  2017.

\bibitem{chen2013distributed}
J.~Chen and A.~H. Sayed,
\newblock ``Distributed pareto optimization via diffusion strategies,''
\newblock {\em IEEE Journal of Selected Topics in Signal Processing}, vol. 7,
  no. 2, pp. 205--220, 2013.

\bibitem{kar2009distributed}
S.~Kar and J.~M.~F. Moura,
\newblock ``Distributed consensus algorithms in sensor networks with imperfect
  communication: Link failures and channel noise,''
\newblock {\em IEEE Transactions on Signal Processing}, vol. 57, no. 1, pp.
  355--369, 2009.

\bibitem{shi2015proximal}
W.~Shi, Q.~Ling, G.~Wu, and W.~Yin,
\newblock ``A proximal gradient algorithm for decentralized composite
  optimization,''
\newblock {\em IEEE Transactions on Signal Processing}, vol. 63, no. 22, pp.
  6013--6023, 2015.

\bibitem{yuan2017exact1}
K.~Yuan, B.~Ying, X.~Zhao, and A.~H. Sayed,
\newblock ``Exact dffusion for distributed optimization and learning -- {Part
  I: Algorithm development},''
\newblock {\em IEEE Trans. Signal Processing}, vol. 67, pp. 724--739, 2018.

\bibitem{nedic2009distributed}
A.~Nedi{\'c} and A.~Ozdaglar,
\newblock ``Distributed subgradient methods for multi-agent optimization,''
\newblock {\em IEEE Transactions on Automatic Control}, vol. 54, no. 1, pp.
  48--61, 2009.

\bibitem{xin2018linear}
R.~Xin and U.~Khan,
\newblock ``A linear algorithm for optimization over directed graphs with
  geometric convergence,''
\newblock {\em IEEE Control Systems Letters}, vol. 2, no. 3, pp. 325--330,
  2018.

\bibitem{chen2015dictionary}
J.~Chen, Z.~J. Towfic, and A.~H. Sayed,
\newblock ``Dictionary learning over distributed models,''
\newblock {\em IEEE Transactions on Signal Processing}, vol. 63, no. 4, pp.
  1001--1016, 2015.

\bibitem{sundhar2012new}
S.~Sundhar, A.~Nedi{\'c}, and V.~V. Veeravalli,
\newblock ``A new class of distributed optimization algorithms: Application to
  regression of distributed data,''
\newblock {\em Optimization Methods and Software}, vol. 27, no. 1, pp. 71--88,
  2012.

\bibitem{mota2012distributed}
J.~F. Mota, J.~M. Xavier, P.~M. Aguiar, and M.~Puschel,
\newblock ``Distributed basis pursuit,''
\newblock {\em IEEE Transactions on Signal Processing}, vol. 60, no. 4, pp.
  1942--1956, 2012.

\bibitem{ying2017diffusion}
B.~Ying and A.~H. Sayed,
\newblock ``Diffusion gradient boosting for networked learning,''
\newblock in {\em Proc. ICASSP}, New Orleans, US, April 2017, pp. 2512--2516.

\bibitem{ying2018exponentially}
B.~Ying, K.~Yuan, and A.~H Sayed,
\newblock ``An exponentially convergent algorithm for learning under
  distributed features,''
\newblock in {\em IEEE Data Science Workshop}, Lausanne, Switzerland, 2018, pp.
  185--189.

\bibitem{ying2018learning}
B.~Ying, K.~Yuan, and A.~H Sayed,
\newblock ``Supervised learning under distributed features,''
\newblock {\em IEEE Transactions on Signal Processing}, vol. 67, pp. 977--992,
  Feb. 2019.

\bibitem{kar2011convergence}
S.~Kar and J.~M.~F. Moura,
\newblock ``Convergence rate analysis of distributed gossip (linear parameter)
  estimation: Fundamental limits and tradeoffs,''
\newblock {\em IEEE Journal of Selected Topics in Signal Processing}, vol. 5,
  no. 4, pp. 674--690, 2011.

\bibitem{shi2015extra}
W.~Shi, Q.~Ling, G.~Wu, and W.~Yin,
\newblock ``{EXTRA}: An exact first-order algorithm for decentralized consensus
  optimization,''
\newblock {\em SIAM Journal on Optimization}, vol. 25, no. 2, pp. 944--966,
  2015.

\bibitem{boyd2006randomized}
S.~Boyd, A.~Ghosh, B.~Prabhakar, and D.~Shah,
\newblock ``Randomized gossip algorithms,''
\newblock {\em IEEE transactions on information theory}, vol. 52, no. 6, pp.
  2508--2530, 2006.

\bibitem{sayed2014adaptive}
A.~H. Sayed,
\newblock ``Adaptive networks,''
\newblock {\em Proceedings of the IEEE}, vol. 102, no. 4, pp. 460--497, April
  2014.

\bibitem{horn1990matrix}
R.~A. Horn and C.~R. Johnson,
\newblock {\em Matrix Analysis},
\newblock Cambridge University Press, 1990.

\bibitem{pillai2005perron}
S.~U. Pillai, T.~Suel, and S.~Cha,
\newblock ``The perron-frobenius theorem: some of its applications,''
\newblock {\em IEEE Signal Processing Magazine}, vol. 22, no. 2, pp. 62--75,
  2005.

\bibitem{freeman2006stability}
R.~A. Freeman, P.~Yang, and K.~M. Lynch,
\newblock ``Stability and convergence properties of dynamic average consensus
  estimators,''
\newblock in {\em Proc. IEEE CDC}, San Diego, CA, 2006, pp. 338--343.

\bibitem{zhu2010discrete}
M.~Zhu and S.~Martinez,
\newblock ``Discrete-time dynamic average consensus,''
\newblock {\em Automatica}, vol. 46, no. 2, pp. 322--329, 2010.

\bibitem{cao2013overview}
Y.~Cao, W.~Yu, W.~Ren, and G.~Chen,
\newblock ``An overview of recent progress in the study of distributed
  multi-agent coordination,''
\newblock {\em IEEE Transactions on Industrial informatics}, vol. 9, no. 1, pp.
  427--438, 2013.

\bibitem{yuan2017exact2}
K.~Yuan, B.~Ying, X.~Zhao, and A.~H. Sayed,
\newblock ``Exact dffusion for distributed optimization and learning -- {Part
  II: Convergence analysis},''
\newblock {\em IEEE Transactions on Signal Processing}, vol. 67, pp. 708--723,
  2018.

\bibitem{tseng2001convergence}
P.~Tseng,
\newblock ``Convergence of a block coordinate descent method for
  nondifferentiable minimization,''
\newblock {\em Journal of optimization theory and applications}, vol. 109, no.
  3, pp. 475--494, 2001.

\bibitem{luo1992convergence}
Z.-Q. Luo and P.~Tseng,
\newblock ``On the convergence of the coordinate descent method for convex
  differentiable minimization,''
\newblock {\em Journal of Optimization Theory and Applications}, vol. 72, no.
  1, pp. 7--35, 1992.

\bibitem{nesterov2012efficiency}
Y.~Nesterov,
\newblock ``Efficiency of coordinate descent methods on huge-scale optimization
  problems,''
\newblock {\em SIAM Journal on Optimization}, vol. 22, no. 2, pp. 341--362,
  2012.

\bibitem{richtarik2014iteration}
P.~Richtarik and M.~Takac,
\newblock ``Iteration complexity of randomized block-coordinate descent methods
  for minimizing a composite function,''
\newblock {\em Mathematical Programming}, vol. 144, no. 1-2, pp. 1--38, 2014.

\bibitem{arablouei2014distributed}
R.~Arablouei, S.~Werner, Y.-F. Huang, and K.~Dogancay,
\newblock ``Distributed least mean-square estimation with partial diffusion,''
\newblock {\em IEEE Transactions on Signal Processing}, vol. 62, no. 2, pp.
  472--484, 2014.

\bibitem{defazio2014saga}
A.~Defazio, F.~Bach, and S.~Lacoste-Julien,
\newblock ``{SAGA}: A fast incremental gradient method with support for
  non-strongly convex composite objectives,''
\newblock in {\em Proc. Advances in Neural Information Processing Systems {\rm
  (NIPS)}}, Montreal, Canada, 2014, pp. 1646--1654.

\bibitem{wang2016coordinate}
C.~Wang, Y.~Zhang, B.~Ying, and A.~H. Sayed,
\newblock ``Coordinate-descent diffusion learning by networked agents,''
\newblock {\em IEEE Transactions on Signal Processing}, vol. 66, no. 2, pp.
  352--367, 2016.

\bibitem{nedic2015distributed}
A.~Nedi{\'c} and A.~Olshevsky,
\newblock ``Distributed optimization over time-varying directed graphs,''
\newblock {\em IEEE Transactions on Automatic Control}, vol. 60, no. 3, pp.
  601--615, 2015.

\bibitem{nedic2016stochastic}
A.~Nedic and A.~Olshevsky,
\newblock ``Stochastic gradient-push for strongly convex functions on
  time-varying directed graphs,''
\newblock {\em IEEE Transactions on Automatic Control}, vol. 61, no. 12, pp.
  3936--3947, 2016.

\bibitem{benezit2010weighted}
F.~B{\'e}n{\'e}zit, V.~Blondel, P.~Thiran, J.~Tsitsiklis, and M.~Vetterli,
\newblock ``Weighted gossip: Distributed averaging using non-doubly stochastic
  matrices,''
\newblock in {\em Proc. International Symposium on Information Theory
  Proceedings}, Austin, Texas, 2010, pp. 1753--1757.

\bibitem{assran2018stochastic}
M.~Assran, N.~Loizou, N.~Ballas, and M.~Rabbat,
\newblock ``Stochastic gradient push for distributed deep learning,''
\newblock {\em arXiv:1811.10792}, 2018.

\bibitem{hu2012robust}
G.~Hu,
\newblock ``Robust consensus tracking of a class of second-order multi-agent
  dynamic systems,''
\newblock {\em Systems \& Control Letters}, vol. 61, no. 1, pp. 134--142, 2012.

\bibitem{di2016next}
P.~D. Lorenzo and G.~Scutari,
\newblock ``Next: In-network nonconvex optimization,''
\newblock {\em IEEE Transactions on Signal and Information Processing over
  Networks}, vol. 2, no. 2, pp. 120--136, 2016.

\bibitem{zouzias2015randomized}
A.~Zouzias and N.~Freris,
\newblock ``Randomized gossip algorithms for solving laplacian systems,''
\newblock in {\em Proc. European Control Conference (ECC)}, 2015, pp.
  1920--1925.

\bibitem{peng2016arock}
Z.~Peng, Y.~Xu, M.~Yan, and W.~Yin,
\newblock ``{ARock}: an algorithmic framework for asynchronous parallel
  coordinate updates,''
\newblock {\em SIAM Journal on Scientific Computing}, vol. 38, no. 5, pp.
  A2851--A2879, 2016.

\bibitem{defazio2014finito}
A.~Defazio, J.~Domke, and T.~S. Caetano,
\newblock ``Finito: A faster, permutable incremental gradient method for big
  data problems.,''
\newblock in {\em Proc. International Conference of Machine Learning {\rm
  (ICML)}}, Beijing, China, 2014, pp. 1125--1133.

\bibitem{hendrikx2018accelerated}
H.~Hendrikx, L.~Massouli{\'e}, and F.~Bach,
\newblock ``Accelerated decentralized optimization with local updates for
  smooth and strongly convex objectives,''
\newblock {\em arXiv preprint arXiv:1810.02660}, Oct. 2018.

\bibitem{kempe2003gossip}
D.~Kempe, A.~Dobra, and J.~Gehrke,
\newblock ``Gossip-based computation of aggregate information,''
\newblock in {\em 44th Annual IEEE Symposium on Foundations of Computer
  Science}, 2003, pp. 482--491.

\bibitem{tsianos2012push}
K.~I. Tsianos, S.~Lawlor, and M.~G. Rabbat,
\newblock ``Push-sum distributed dual averaging for convex optimization,''
\newblock in {\em 51st IEEE Conference on Decision and Control (CDC)}, Maui,
  HI, 2012, pp. 5453--5458.

\bibitem{dominguez2011distributed}
A.~Dominguez-Garcia and C.~Hadjicostis,
\newblock ``Distributed strategies for average consensus in directed graphs,''
\newblock in {\em IEEE Conference on Decision and Control and European Control
  Conference}, Orlando, Florida, 2011, pp. 2124--2129.

\bibitem{notarnicola2018distributed}
I.~Notarnicola, Y.~Sun, G.~Scutari, and G.~Notarstefano,
\newblock ``Distributed big-data optimization via block-iterative gradient
  tracking,''
\newblock {\em available at arXiv:1808.07252}, Aug. 2018.

\bibitem{simonetto2017decentralized}
A.~Simonetto, A.~Koppel, A.~Mokhtari, G.~Leus, and A.~Ribeiro,
\newblock ``Decentralized prediction-correction methods for networked
  time-varying convex optimization,''
\newblock {\em IEEE Transactions on Automatic Control}, vol. 62, no. 11, pp.
  5724--5738, 2017.

\bibitem{ling2013decentralized}
Q.~Ling and A.~Ribeiro,
\newblock ``Decentralized dynamic optimization through the alternating
  direction method of multipliers,''
\newblock {\em IEEE Transactions on Signal Processing}, vol. 62, no. 5, pp.
  1185--1197, 2013.

\bibitem{olfati2007distributed}
R.~Olfati-Saber,
\newblock ``Distributed kalman filtering for sensor networks,''
\newblock in {\em 46th IEEE Conference on Decision and Control}, New Orleans,
  LA, 2007, pp. 5492--5498.

\bibitem{olfati2005distributed}
R.~Olfati-Saber,
\newblock ``Distributed kalman filter with embedded consensus filters,''
\newblock in {\em Proceedings of the 44th IEEE Conference on Decision and
  Control}, Seville, Spain, 2005, pp. 8179--8184.

\bibitem{cattivelli2010diffusion}
F.~S. Cattivelli and A.~H. Sayed,
\newblock ``Diffusion strategies for distributed kalman filtering and
  smoothing,''
\newblock {\em IEEE Transactions on automatic control}, vol. 55, no. 9, pp.
  2069--2084, 2010.

\bibitem{xu2015augmented}
J.~Xu, S.~Zhu, Y.~C. Soh, and L.~Xie,
\newblock ``Augmented distributed gradient methods for multi-agent optimization
  under uncoordinated constant stepsizes,''
\newblock in {\em 54th IEEE Conference on Decision and Control (CDC)}, 2015,
  pp. 2055--2060.

\bibitem{xi2017add}
C.~Xi, R.~Xin, and U.~A Khan,
\newblock ``Add-opt: Accelerated distributed directed optimization,''
\newblock {\em IEEE Transactions on Automatic Control}, vol. 63, no. 5, pp.
  1329--1339, 2017.

\bibitem{qu2017harnessing}
G.~Qu and N.~Li,
\newblock ``Harnessing smoothness to accelerate distributed optimization,''
\newblock {\em IEEE Transactions on Control of Network Systems}, vol. 5, no. 3,
  pp. 1245--1260, 2017.

\bibitem{ying2018performance}
B.~Ying and A.~H. Sayed,
\newblock ``Performance limits of stochastic sub-gradient learning, part ii:
  Multi-agent case,''
\newblock {\em Signal Processing}, vol. 144, pp. 253--264, 2018.

\bibitem{vlaski2016diffusion}
S.~Vlaski, L.~Vandenberghe, and A.~H. Sayed,
\newblock ``Diffusion stochastic optimization with non-smooth regularizers,''
\newblock in {\em Proc. ICASSP}, Shanghai, China, Mar. 2016, pp. 4149--4153.

\bibitem{tu2012diffusion}
S.-Y. Tu and A.~H. Sayed,
\newblock ``Diffusion strategies outperform consensus strategies for
  distributed estimation over adaptive networks,''
\newblock {\em IEEE Transactions on Signal Processing}, vol. 60, no. 12, pp.
  6217--6234, 2012.

\bibitem{li2017decentralized}
Z.~Li, W.~Shi, and M.~Yan,
\newblock ``A decentralized proximal-gradient method with network independent
  step-sizes and separated convergence rates,''
\newblock {\em To appear in IEEE Transaction on Signal Processing. Also
  available at arXiv:1704.07807}, 2017.

\bibitem{pu2018push}
S.~Pu, W.~Shi, J.~Xu, and A.~Nedic,
\newblock ``A push-pull gradient method for distributed optimization in
  networks,''
\newblock {\em available at arXiv:1810.06653}, Nov. 2018.

\bibitem{tahbaz2008necessary}
A.~Tahbaz-Salehi and A.~Jadbabaie,
\newblock ``A necessary and sufficient condition for consensus over random
  networks,''
\newblock {\em IEEE Transactions on Automatic Control}, vol. 53, no. 3, pp.
  791--795, 2008.

\bibitem{bremaud2013markov}
P.~Br{\'e}maud,
\newblock {\em Markov Chains: Gibbs Fields, Monte Carlo Simulation, and
  Queues},
\newblock Springer, 2013.

\bibitem{Kolmogorov1991elements}
A.~N. Kolmogorov and S.~Fomin,
\newblock {\em Elements of the Theory of Functions and Functional Analysis},
\newblock Courier Corporation, 1999.

\bibitem{boyd2004convex}
S.~Boyd and L.~Vandenberghe,
\newblock {\em Convex Optimization},
\newblock Cambridge University Press, 2004.

\bibitem{mao2018walkman}
X.~Mao, K.~Yuan, Y.~Hu, Y.~Gu, A.~H. Sayed, and W.~Yin,
\newblock ``Walkman: A communication-efficient random-walk algorithm for
  decentralized optimization,''
\newblock {\em Available at arXiv:1804.06568}, April 2018.

\end{thebibliography}

\end{document}